\documentclass[aip, jcp, final, reprint]{revtex4-1}

\usepackage{amsmath}
\usepackage{amsfonts}
\usepackage{amssymb}
\usepackage{gensymb}
\usepackage{graphicx}
\usepackage[colorlinks=true, linkcolor=black, urlcolor=blue, citecolor=black, anchorcolor=black]{hyperref}

\usepackage{enumitem}
\setlist{nosep}

\def\app#1#2{%
	\mathrel{%
		\setbox0=\hbox{$#1\sim$}%
		\setbox2=\hbox{%
			\rlap{\hbox{$#1\propto$}}%
			\lower1.1\ht0\box0%
		}%
		\raise0.25\ht2\box2%
	}%
}

\graphicspath{{"figures/"}}

\usepackage{xcolor}
\usepackage{ulem}
\newcommand{\del}[1]{\textcolor{red}{\sout{#1}}}

\begin{document}

\title{Multidimensional Triple Sum-Frequency Spectroscopy of MoS\textsubscript{2} and Comparisons with Absorption and Second Harmonic Generation Spectroscopies}
\author{Darien J. Morrow}
\author{Daniel D. Kohler}
\author{Kyle J. Czech}
\author{John C. Wright}
\email{wright@chem.wisc.edu}
\affiliation{Department of Chemistry,
	University of Wisconsin--Madison,
	1101 University Ave, 
	Madison, WI 53706, United States}
%\keywords{THG, TSF, CMDS, MoS2, TMDC}

\date{\today} 

\begin{abstract}
	Triple sum-frequency (TSF) spectroscopy is a recently-developed methodology that enables collection of multidimensional spectra by resonantly exciting multiple quantum coherences of vibrational and electronic states. 
	This work reports the first application of TSF to the electronic states of semiconductors. 
	Two independently tunable ultrafast pulses excite the A, B, and C features of a MoS\textsubscript{2} thin film. 
	The measured TSF spectrum differs markedly from absorption and second harmonic generation spectra. 
	The differences arise because of the relative importance of transition moments and the joint density of states. 
	We develop a simple model and globally fit the absorption and harmonic generation spectra to extract the joint density of states and the transition moments from these spectra.
	Our results validate previous assignments of the C feature to a large joint density of states created by band nesting. 	
\end{abstract}

\maketitle

%\section{Introduction}

Coherent multidimensional spectroscopy (CMDS) is a useful tool for exploring the rich many-body physics of semiconductors.\cite{Cundiff_2008, Moody_Cundiff_2017} 
A new CMDS methodology is Triple Sum Frequency (TSF) spectroscopy.
TSF is the non-degenerate analog of third harmonic generation (THG), and the four-wave mixing extension of three-wave mixing processes like sum-frequency generation and second harmonic generation (SHG).\cite{NeffMallon_Wright_2017} 
TSF uses independently tunable ultrafast pulses to create coherences at increasingly higher frequencies while discriminating against transient populations. 
Scanning the multiple input pulse frequencies enables collection of a multidimensional spectrum.
Cross peaks in the spectrum identify the dipole coupling between states.
TSF has studied vibrational and electronic coupling in molecules.\cite{Boyle_Wright_2013, Boyle_Wright_2013_001, Boyle_Wright_2014, Grechko_Bonn_2018} 
This work reports the first TSF spectroscopy of a semiconductor. 

We investigate a polycrystalline MoS\textsubscript{2} thin film.  
Transition metal dichalcogenides (TMDCs), such as MoS\textsubscript{2}, are layered semiconductors whose indirect bandgaps become direct in the monolayer limit.\cite{Mak_Heinz_2010, Ellis_Scuseria_2011}
TMDCs exhibit strong spin-orbit coupling, high charge mobility, and have novel photonic capabilities.\cite{Wang_Strano_2012, Xia_Ramasubramaniam_2014, Mak_Shan_2016} 
The optical spectrum of MoS\textsubscript{2} is dominated by three features: A ($\hbar\omega_\text{A} \approx 1.8 \text{ eV}$), B ($\hbar\omega_\text{B} \approx 1.95 \text{ eV}$), and C ($\hbar\omega_\text{C} \approx 2.7 \text{ eV}$).\cite{MolinaSanchez_Wirtz_2013, Li_Heinz_2014}
A and B originate from high binding energy excitonic transitions between spin-orbit split bands.\cite{MolinaSanchez_Wirtz_2013, Qiu_Louie_2013, He_Shan_2014, Saigal_Ghosh_2016, Kopaczek_Kudrawiec_2016}
The stronger C feature is predicted to arise from a large joint density of states (JDOS) due to band nesting across a large section of the Brillouin zone (BZ).\cite{Britnell_Novoselov_2013, Carvalho_Neto_2013, Jeong_Cho_2018, Bieniek_Hawrylak_2018}
As of yet, no direct, experimental verification of the large JDOS defining the C feature has been accomplished.
In this work, we demonstrate how first, second, and third order spectroscopies can be used together to determine whether the prominence of a feature is due to a large transition dipole or a large transition degeneracy.

The spectroscopies considered here can be understood in the electric dipole approximation using perturbation theory.\cite{Boyd_2008, Bloembergen_Shen_1964}
Briefly, an electric field ($E$) drives a polarization ($P$) in the material.
The polarization is related to an oscillating coherence between two states.
The polarization is expressed as an expansion in electric field and susceptibility ($\chi$) order. 
Absorption, SHG, and TSF (THG) depend on $\chi^{(1)}$, $\chi^{(2)}$, and $\chi^{(3)}$, respectively.
Absorption is proportional to $\text{Im}\left[\chi^{(1)}\right]$.
For very thin films (no interference or velocity-mismatch effects), SHG and TSF intensities are proportional to $\left| \chi^{(2)}\right|^2$ and $\left| \chi^{(3)}\right|^2$, respectively.
All of the discussed spectroscopies detect state coupling through resonant enhancement.
When the driving field is resonant with an interstate transition, $\chi$ is enhanced and the output intensity increases.
In crystalline systems, interstate transitions are also subject to momentum conservation, which typically restricts interstate coupling to vertical transitions within the Brillouin zone (i.e. direct transitions).

Our work expands upon the extensive body of SHG and THG work
on TMDCs\cite{Li_Heinz_2013, Malard_dePaula_2013, Kumar_Zhao_2013, Wang_Zhao_2013, Trolle_Pedersen_2014, Gruning_Attaccalite_2014, Clark_Kim_2014, Trolle_Pedersen_2015, Wang_Urbaszek_2015, Sun_Chan_2016, Karvonen_Sun_2017, Shearer_Jin_2017, Glazov_Urbaszek_2017, Balla_McCloskey_2018} by exploring the multidimensional frequency response.
TSF and other CMDS spectroscopies that scan multiple driving field frequencies can identify multiresonant enhancement, whereby the driving fields resonate with more than one interstate transition.\cite{Wright_2011}
Multiresonance selectively enhances coupled transitions and decongests dense spectra.
In crystalline materials, multiresonant excitation is also subject to the momentum conservation mentioned earlier, so TSF can isolate multiple transitions from singular points in the Brillouin zone.
Because of this selectivity, TSF is a potential method for mapping out band dispersion in crystals.
With three independently tunable lasers, TSF can couple up to four states at a specific $\mathbf{k}$-point together, so TSF could measure the dispersion of up to four bands.
The present work is a step towards the goal of momentum-selective CMDS.

%\section{Experimental}

Nonlinear measurements of thin films are often complicated by non-resonant substrate contributions which are mitigated by measuring the coherent output in the reflected instead of transmitted direction.\cite{Volkmer_Xie_2001, Czech_Wright_2015, Morrow_Wright_2017, Handali_Wright_2018, Honold_Tu_1988} 
For our experiment, TSF measured in the reflected direction has an effective penetration depth of  $\sim \lambda_{\textrm{fundamental}}/12 \sim 100 \: \textrm{nm}$; this small sampling length allows the resonant response from the thin film to be orders of magnitude more intense than the non-resonant response (see SI for more discussion).
We prepared a 10 nm thick MoS\textsubscript{2} thin film by first evaporating Mo onto the fused silica substrate followed by sulfidation of the Mo.\cite{Czech_Wright_2015, Laskar_Rajan_2013}\del{.} 
Our sample substrate is a fused silica prism so that back-reflected, non-resonant TSF exits the substrate traveling parallel to the desired TSF signal but shifted spatially.
Sample geometry, synthesis details, AFM measurements to determine film thickness, and a Raman spectrum are present in the SI.

For our TSF measurements, an ultrafast oscillator seeds a regenerative amplifier, creating pulses centered at 1.55 eV with a 1 kHz repetition rate. 
These pulses pump two optical parametric amplifiers (OPAs) which create tunable pulses of light from $\sim0.5$ to $\sim1\text{ eV}$ with spectral width on the amplitude level of $\text{FWHM} \approx 46 \:\textrm{meV}$.
The two beams with frequencies $\omega_1$ \& $\omega_2$ and wave vectors $\vec{k}_1$ \&  $\vec{k}_2$ are focused onto the sample.
All beams are linearly polarized (S relative to sample) and coincident in time.
The spatially coherent output with wave vector $-\left(\vec{k}_1 + 2\vec{k}_2\right)$ is isolated with an aperture (the negative sign\del{s} correspond to the reflective direction), focused into a monochromator, and detected with a photomultiplier tube. 
The TSF intensity is linear in $\omega_1$ fluence and quadratic in $\omega_2$ fluence (see the SI for details).
The measured MoS\textsubscript{2} TSF spectrum is normalized by the measured TSF spectrum of the fused silica substrate in order to account for spectrally-dependent OPA output intensities and detector responsivity.
The SI contains additional experimental and calibration details.
All raw data, workup scripts, and simulation scripts used in the creation of this work are permissively licensed and publicly available for reuse.\cite{OSF}
Our acquisition,\cite{PyCMDS} workup,\cite{WrightTools} and modeling software\cite{OSF} are built on top of the open source, publicly available Scientific Python ecosystem.\cite{Jones_2001, vanderWalt_Varoquaux_2011, Hunter_2007}

%\section{Experimental Results}

\begin{figure}[!htbp]
	\includegraphics[width=0.5 \textwidth]{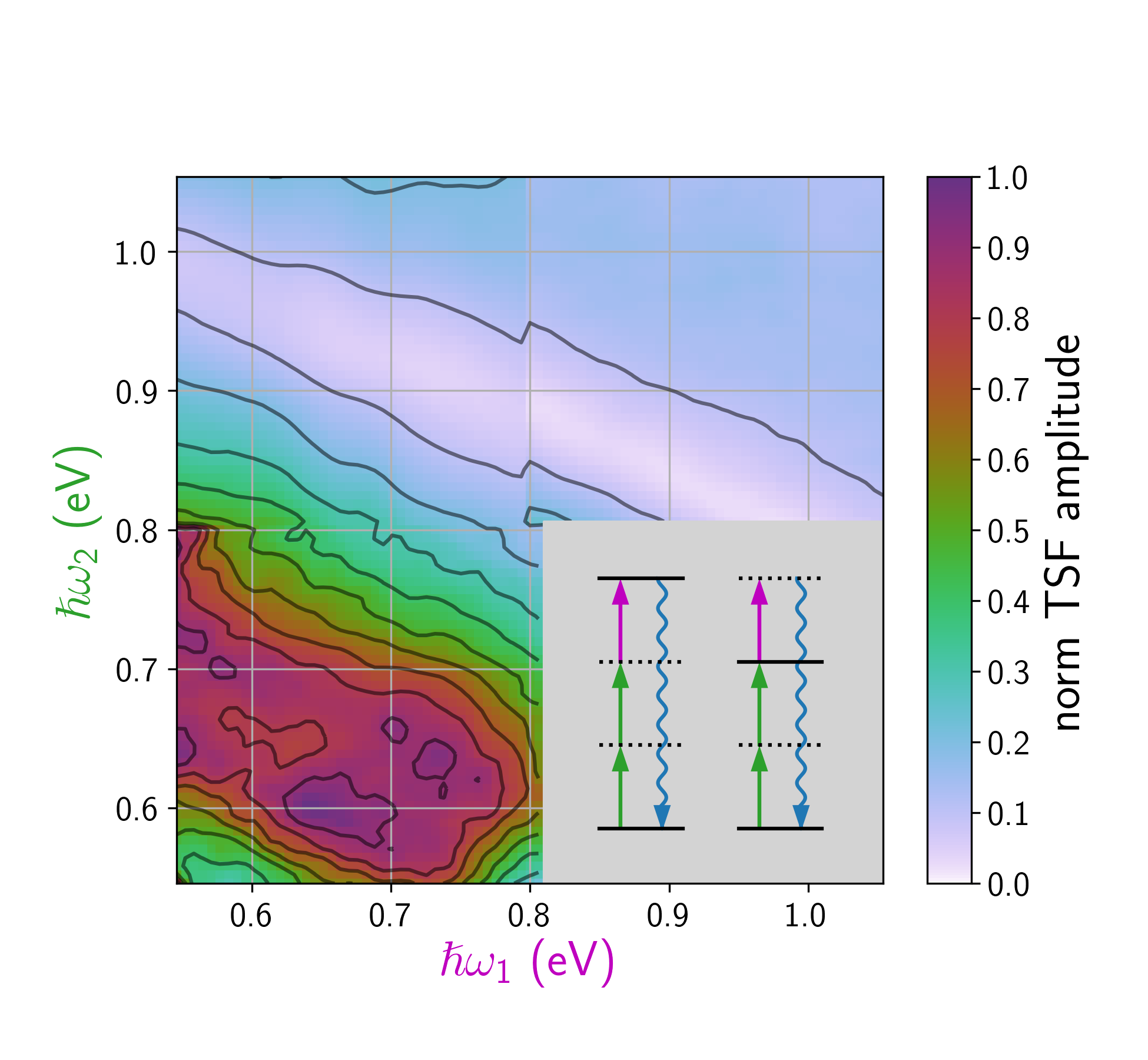}
	\caption{\label{F:TSF}
		Normalized 2D TSF spectra of MoS\textsubscript{2}. 
		Inset diagrams processes where the third interaction is resonant with a state (left) and when the second interaction is resonant with a state (right). 
		The measured output is represented by a wavy downward arrow. 
		Note, the gray area of the inset was not experimentally explored.
	}
\end{figure}

Our main experimental result is a 2D TSF spectrum of a MoS\textsubscript{2} thin film (\autoref{F:TSF}).
The TSF spectrum of MoS\textsubscript{2} has a simple structure, with all features running parallel to a line with slope of -1/2. 
This single variable dependence of our output intensity implies there is negligible multiresonant enhancement within our spectral window.
Our spectral window specifically rules out optical coupling between the A and C features.
If the A and C features were coupled we would see a peak that depends on two frequency conditions: $\hbar\omega_2 = \hbar \omega_\text{A}/2 \approx 0.9 \text{ eV}$ and $\hbar\omega_1 = \hbar\left(\omega_\text{C} - \omega_\text{A}\right) \approx 0.9 \text{ eV}$.
Note that A and C features are believed to originate from different regions of the Brillouin zone, so resonant enhancement is not expected.

The slope of -1/2 implies that only the output frequency, $\omega_1 + 2 \omega_2$, determines the resonant enhancement.
The energy ladder diagram for such a resonant enhancement is shown in the \autoref{F:TSF} inset (left).
Peaks at output colors $\sim$1.85 \& $\sim$2.0 eV are roughly consistent with three photon resonances with the A and B features, while a trough is present for output colors close to the C feature ($\sim 2.7$ eV).  
Curiously, we do not see contributions from two-photon resonances (energy level diagram \autoref{F:TSF} inset, right), even when $2\omega_2$ or $\omega_1+\omega_2$ traces over the A or B features.
The two-photon resonances would manifest as horizontal or anti-diagonal (slope of -1) features in \autoref{F:TSF}.
The lack of two-photon resonances was surprising to us given the large two-photon absorption cross-section of MoS\textsubscript{2}.\cite{Zhang_Wang_2015}
We address this observation later.

\begin{figure}[!htbp]
	\includegraphics[width=0.5 \textwidth]{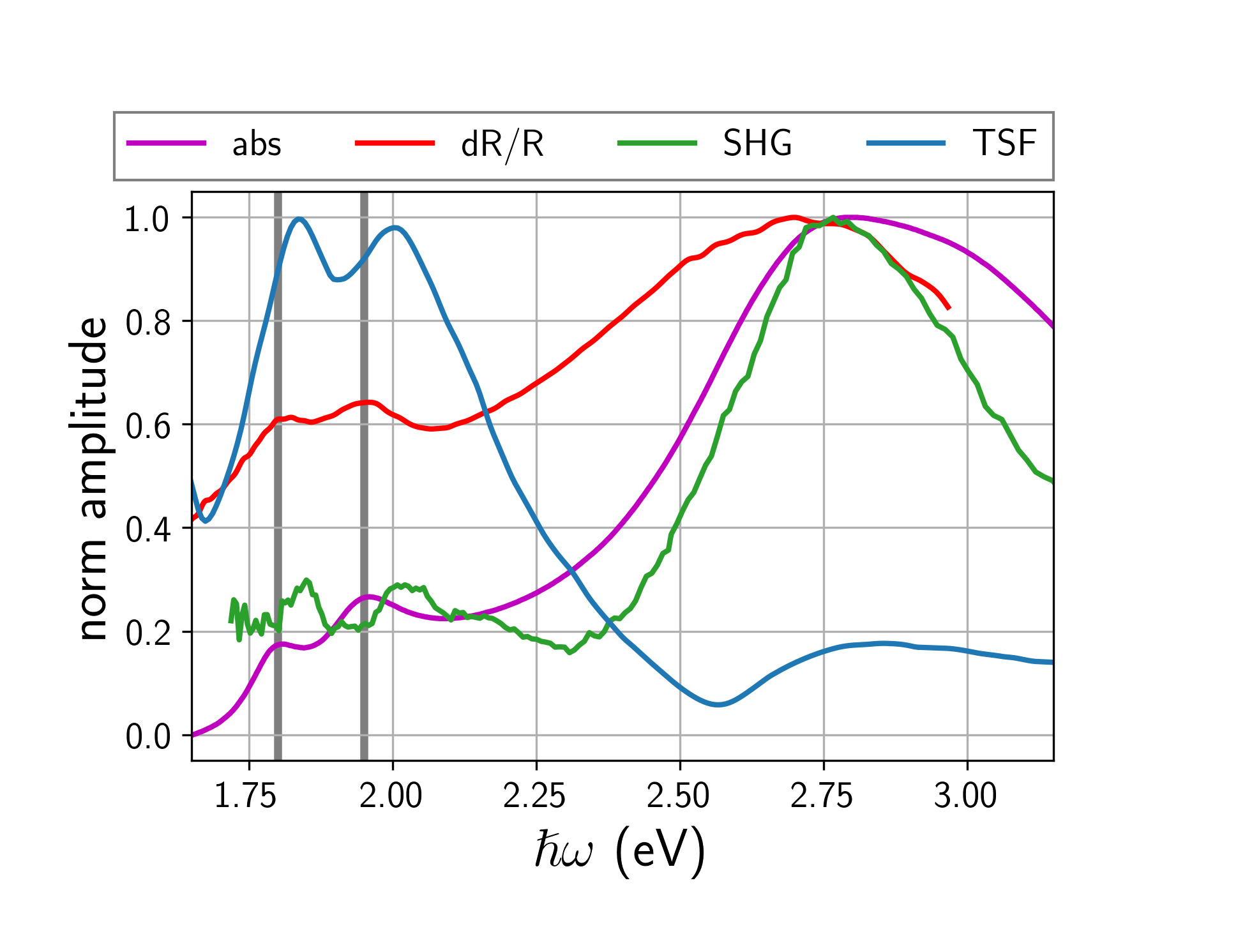}
	\caption{\label{F:1D}
		Normalized amplitude 1D spectra of MoS\textsubscript{2} thin films: absorption\cite{Czech_Wright_2015}, SHG\cite{Trolle_Pedersen_2015}, reflection contrast (dR/R), and TSF (THG). The TSF spectrum is derived from \autoref{F:TSF} as detailed in the main text. Vertical gray bars are guides to the eye set at 1.80 and 1.95 eV to demonstrate how SHG and TSF features are blue shifted from absorption and dR/R spectral peaks. 
	}
\end{figure}

Since the dominant spectral features in \autoref{F:TSF} depend only on output color, we can generate a 1D THG spectrum by plotting the mean TSF amplitude for each output color.
The THG spectrum is compared with other techniques in \autoref{F:1D}. 
Due to the unconventional prism substrate, we were unable to acquire an absorption spectrum of the sample, but we did acquire a reflection contrast spectrum shown in \autoref{F:1D}.\cite{McIntyre_Aspens_1971, Mak_Heinz_2008}
The absorption spectrum presented in \textcite{Czech_Wright_2015} and \autoref{F:1D} is of a sample prepared with similar conditions as our sample, but on a flat window substrate.
The A and B feature peaks of the THG spectrum are blue-shifted compared to the absorption spectrum.
\textcite{Wang_Zhao_2013} observed a similar blue shift when they measured the THG spectrum of MoS\textsubscript{2} around the A and B features.
The C feature is dominant in the SHG\cite{Trolle_Pedersen_2015} and absorption spectrum\cite{Czech_Wright_2015} while the A and B features are dominant in the THG spectrum. 
This observation is the main motivation for our analysis.

%\section{Theory}

To explain why C dominates the absorption and SHG spectra but not the THG spectrum, we develop a simple model.
To our knowledge, a unified model for comparing large dynamic-range absorption, SHG, and THG spectra of a semiconductor does not exist. 
Notable headway has been made to calculate SHG spectra\cite{Sipe_Shkrebtii_2000} and write closed equations of motion\cite{Axt_Mukamel_1998} for semiconductors, but simple formalisms are lacking.
Most simple formalisms (c.f. refs.\cite{Peychambarian_Mysyrowicz_1993, Dresselhaus_2018}) approximate the dipole as a constant with respect to both transition energy and lattice momentum.
This constant dipole approximation breaks down above the bandgap, where the lattice momentum of transitions is significantly different from that of the bandgap transitions.
As the lattice momentum changes, Bloch waves alter their bonding symmetries and intralattice character (cf. \textcite{Padilha_VandeWalle_2014}), which alters the transition dipole moments.
Since the C feature is believed to originate from a different region of the Brillouin Zone than the bandedge transitions, our theory requires both the dipole moment and JDOS to vary across our spectral range.

We develop our simple model by expanding the typical linear response formalism for direct transitions of semiconductors to include sum-frequency processes.  
In the case of $\chi^{(1)}$, simple theories exist for expanding the single oscillator case to bulk conditions.  
For more than one oscillator, the total susceptibility is the sum of individual susceptibilities.
For a semiconductor system, this is a summation over all wave-vectors, $\left\{\mathbf{k}\right\}$ such that $\mathbf{k}\in\text{BZ}$:
\begin{gather} 
	\chi^{\left(1\right)} \left(-\omega_1, \omega_1\right) = 
		\sum_{a, g} \sum_{\mathbf{k}}
		\frac{\mu_{ga\mathbf{k}}^2}{
			\Delta_{ga\mathbf{k}}^{1}}, \label{E:chi1}
\end{gather}
where $\Delta_{ga\mathbf{k}}^{1} = \omega_{ag\mathbf{k}} - \omega_1 -i\Gamma$ and $\Gamma$ is a damping rate which accounts for the finite width of the optical transitions. 
It is common to replace the summation with a transition energy distribution function between conduction band $x$ and valence band $y$, $J_{xy}(E)$,\footnote{
	\autoref{E:chi1}, and all further theory developed, neglect indirect transitions. 
	We find this a reasonable assumption since our multidimensional spectrum exhibited no cross-peaks between the K-point features (A and B) and the C band.
}
such that
\begin{equation}\label{E:chi1_cv}
	\chi^{(1)}(-\omega_1, \omega_1) \propto \int \frac{\text{d}E}{E-\hbar\omega_1 -i\hbar\Gamma} \sum_{x,y} J_{xy}(E) \mu_{xy}(E)^2.
\end{equation} 
In crystals, $J_{xy}(E)$ is the JDOS.
Note the dependence of the susceptibility on both $J_{xy}(E)$ and $\mu_{xy}(E)^2$. 
In the case that either the JDOS or the transition dipole are constant, spectroscopy techniques can be used to locate excitonic transitions or critical points in the JDOS.
If the JDOS and transition dipole both vary, then traditional techniques fail.

The THG and SHG responses take the form of:
\begin{gather}
\chi^{\left(2\right)} \left(-\omega_{21}, \omega_1, \omega_2\right) = 
	\mathcal{P} \sum_{b, a, g}\sum_{\mathbf{k}} 
		\frac{\mu_{gb\mathbf{k}} \mu_{ba\mathbf{k}} \mu_{ag\mathbf{k}}}{
			\Delta_{gb\mathbf{k}}^{12}
			\Delta_{ga\mathbf{k}}^{1}} \label{E:chi2}\\
\chi^{\left(3\right)} \left(-\omega_{321}, \omega_1, \omega_2, \omega_3\right) = 
	\mathcal{P} \sum_{c, b, a, g}\sum_{\mathbf{k}} 
		\frac{\mu_{gc\mathbf{k}} \mu_{cb\mathbf{k}} \mu_{ba\mathbf{k}} \mu_{ag\mathbf{k}}}{
			\Delta_{gc\mathbf{k}}^{123}
			\Delta_{gb\mathbf{k}}^{12}
			\Delta_{ga\mathbf{k}}^{1}}, \label{E:chi3}
\end{gather}
where $c,b,a$, and $g$ are bands of the semiconductor.  
We have defined $\omega_{21} \equiv \omega_2 + \omega_1$ and $\omega_{321}\equiv \omega_3 + \omega_2 + \omega_1$.
$\mathcal{P}$ is a permutation operator which accounts for all combinations of field-matter interactions.
The additional detuning factors are defined by $\Delta_{gc\mathbf{k}}^{123} \equiv \omega_{cg\mathbf{k}}-\omega_{321} -i\Gamma$ and $\Delta_{gb\mathbf{k}}^{12} \equiv \omega_{bg\mathbf{k}} -\omega_{21} -i\Gamma$ in which $\omega_{ab}$ is the frequency difference between bands $a$ and $b$ at point $\mathbf{k}$ in the BZ. 
The JDOS formalism employed in \autoref{E:chi1_cv} can be abstracted to describe $\chi^{(2)}$ and $\chi^{(3)}$ with the introduction of multidimensional joint density functions.
These joint densities depend not just on the energy difference between the initial and final states, but also on the energy differences between the intermediate states reached during the sum-frequency process.
See the SI for further details. 

To simulate the spectra, the sum over bands in \autoref{E:chi1_cv}-\autoref{E:chi3} is truncated at three total bands:  the valence band, $v$, the conduction band, $c$, and a third higher-energy band, $b$, (note: bands $b$ and $c$ are not to be confused with the B and C absorption spectrum features).
The SHG and THG spectra are measured close to the direct bandgap, so transitions between $c$ and $v$ are key to describing the response.
The $b$ band is taken to be a much higher energy (6 eV) than the valence band.
We define the transition strength of low-lying states ($c$ and $v$) to this nondescript high-lying band with the parameter $\mu_{\text{NR}} (\hbar \omega)$.
We note that $\mu_{\text{NR}}$ is not formally a dipole, but instead contains all non-resonant transition factors involving band $b$ (dipoles and degeneracies between $c$ and $b$ or $v$ and $b$).
While this is a improper parameterization of the actual band structure above the conduction band and below the valence band, 
its inclusion is crucial for reproducing details of our spectra, and the parameters offer insight into the role of virtual states in sum-frequency spectroscopies.

With this framework, we can now reason why THG, SHG, and absorption measurements are complementary for distinguishing degeneracy and dipole moments.  
The strength of absorption is proportional to $\mu_{cv}^2$ and $J_{cv}$ (\autoref{E:chi1_cv}).
SHG signals will be due to the sequence $v \rightarrow b \rightarrow c$, which informs on the non-resonant band $b$.
THG has sequences such as $v \rightarrow c \rightarrow v \rightarrow c$, which scale as $\mu_{cv}^4$ but are still linear in $J_{cv}$.  
On the other hand, THG can depend on state $b$ and consequently depends on the same non-resonant features of SHG. 
THG and absorption give different scalings for transitions between $c$ and $v$, but SHG is also needed to constrain the non-resonant transitions of THG involving band $b$.

%\section{Modeling}

\begin{figure*}[!htbp]
	\includegraphics[width=\textwidth]{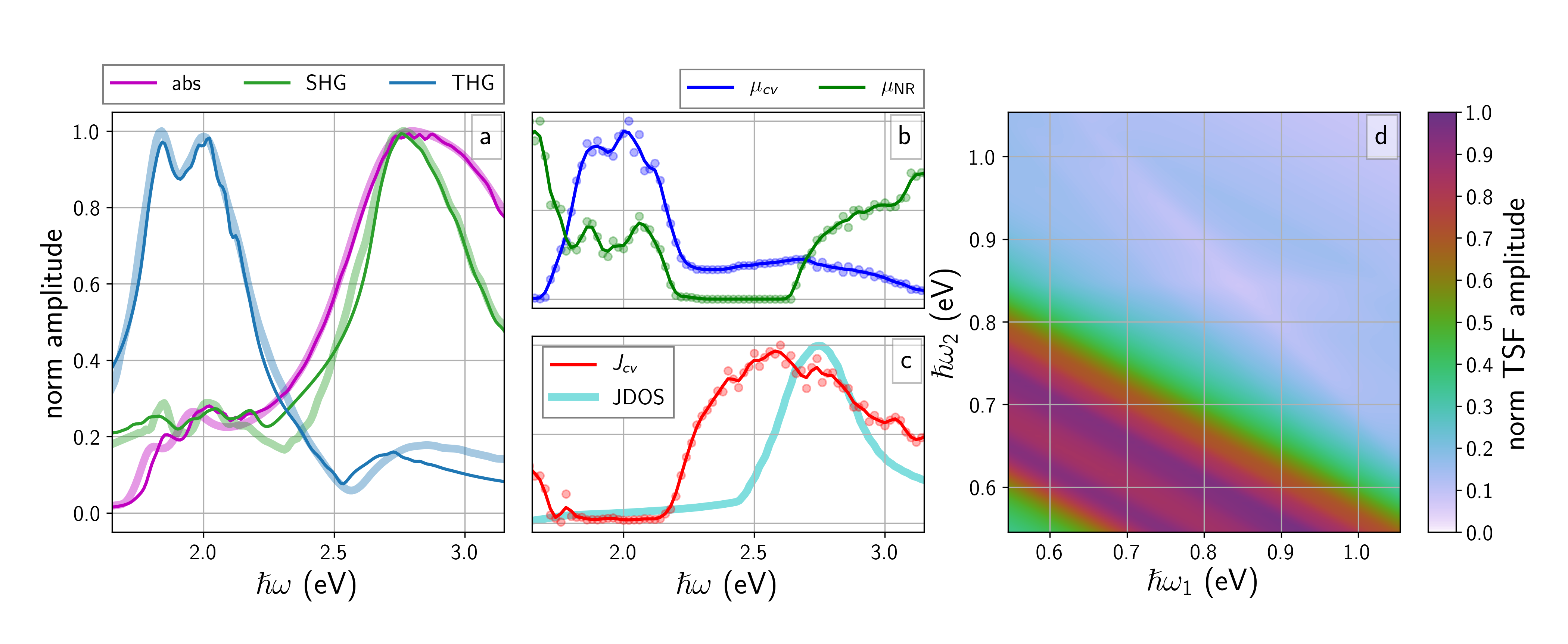}
	\caption{\label{F:sims}
		Variational model of optical spectroscopies.
		(a) normalized comparison of experiment (thick, translucent lines, absorption\cite{Czech_Wright_2015} and SHG\cite{Trolle_Pedersen_2015}) and model (thin lines).
		(b) normalized model dipole parameters in experimentally explored range.
		(c) normalized model density, $J_{cv}$, and tight-binding optical JDOS from \textcite{Bieniek_Hawrylak_2018}.
		(d) normalized TSF spectrum as predicted by our model as fit to 1D experiments.} 
\end{figure*}

\autoref{F:sims} summarizes the fit of our model to experiment. 
Our simulation uses a discrete set of transition energies to approximate the integral of \autoref{E:chi1_cv}.
We employ 140 discrete energies spaced 20 meV apart with $\hbar\Gamma = 20\text{ meV}$, $\hbar\omega_{b_j} = 6 \text{ eV}$, and $\hbar\Gamma_b=500\text{ meV}$.
Our strategy of a discretized set of transition energies is similar to the constrained variational analysis that is often employed to relate a material's reflection spectrum to its absorption or dielectric spectra.\cite{Kuzmenko_2005, Li_Heinz_2014, Sie_Gedik_2015} 
As a function of transition energy, the model extracts the dipole strength of the $c\leftrightarrow v$ transitions, $\mu_{cv}$; a weighting factor for transitions involving the non-resonant state, $\mu_\text{NR}$; and the transition degeneracy, $J_{cv}$.
See the SI for further modeling details.   

\autoref{F:sims}a shows qualitative agreement between the model and experiment.
Our model does not assume a functional form for the JDOS or dipole spectra, so we can compare the response from an excitonic transition to that of an interband transition.
The fitted parameters are shown in \autoref{F:sims}b and \autoref{F:sims}c; note that both $\mu_{cv}$ and $\mu_\text{NR}$ are peaked near the A and B features, while $J_{cv}$ is minimized. 
As $\hbar\omega$ increases from the B feature, $J_{cv}$ drastically increases while $\mu_{cv}$ and $\mu_\text{NR}$ both decrease---the increase in $J_{cv}$ is analogous to the large JDOS attributed to band nesting by recent workers.\cite{Britnell_Novoselov_2013, Carvalho_Neto_2013, Jeong_Cho_2018} 
For comparison, in \autoref{F:sims}c we plot the JDOS as recently calculated by \textcite{Bieniek_Hawrylak_2018} for monolayer MoS\textsubscript{2} within their tight-binding model.
Both the extracted $J_{cv}$ and the tight-binding, monolayer JDOS have a small value near the A and B features but form a peaked structure near the C feature.
Because absorption, SHG, and THG spectra all scale linearly with $J_{cv}$ but differently with transition dipole strength, the extracted structure of $J_{cv}$ and $\mu_{cv}$ explains the glaring disparities between THG and the other two spectra. 
Our fitting procedure convincingly reproduces the large degeneracy of the C feature due to band-nesting.

To the red of the A feature, the JDOS and $\mu_\text{NR}$ increase while $\mu_{cv}$ decays to zero. 
We attribute this behavior to an artifact of our finite spectral range.  
Variational approaches are known to have difficulty with the edges of spectra.\cite{Roessler_1965}

Though our fitting procedure examined only harmonic generation, our fits also explain the notable lack of two-photon resonances in the 2D TSF spectrum.
\autoref{F:sims}d shows a 2D TSF spectrum simulated from our fit parameters.  
The TSF features produced by our model are primarily three photon resonances which lie parallel to lines of constant output color.
Some features from two-photon resonances (e.g. $v \rightarrow b \rightarrow c \rightarrow b$), such as the trough over the anti-diagonal line $\hbar(\omega_1 + \omega_2) = 1.7 \text{ eV}$ are visible but minor.  
The two-photon resonances that are prominent in the SHG spectra are suppressed because transitions involving $b$ are severely detuned so sequences like $v \overset{\mu_{cv}}{\rightarrow} c \overset{\mu_{cv}}{\rightarrow} v \overset{\mu_{cv}}{\rightarrow} c$ dominate the output over sequences like $v \overset{\mu_{\text{NR}}}{\rightarrow} b \overset{\mu_{\text{NR}}}{\rightarrow} c \overset{\mu_{\text{NR}}}{\rightarrow} b$.    
In the model, below the line $\hbar(\omega_1 + \omega_2) = 1.7 \text{ eV}$, the dominant features come from coherence pathways like $v \rightarrow c \rightarrow b \rightarrow c$ or $v \rightarrow c \rightarrow v \rightarrow c$, in which the third excitation is resonant.

Our model fails to capture some features of the three optical measurements.
For instance, the model shifts the position of A and B absorption features and undershoots the THG spectrum at energies above 2.7 eV.
The model falsely attributes the nature of the deep trough seen in \autoref{F:TSF} which runs along $\hbar(\omega_1 + 2\omega_2) \approx 2.56 \text{ eV}$.
Specifically, \autoref{F:sims}d shows the dip of the simulated THG spectrum to be due to a two photon resonance running along $\hbar(\omega_1 + \omega_2) \approx 1.7 \text{ eV}$; this resonance is not seen in \autoref{F:TSF} which exclusively exhibits three photon resonances. 
A more complete fit would take into account our full 2D spectrum in order to distinguish between pathways with resonance enhancement from the second and third interaction. 
A more careful treatment of non-resonant transitions ($\mu_{\text{NR}}$) may suppress these pathways. 
For instance, our model extracts a deep trough in $\mu_{\text{NR}}$ around 2.2-2.6 eV, yet we do expect the non-resonant transition strength to have a strong dependence on our output color.
Despite these shortcomings, our approach provides a robust characterization that informs on the interplay of dipole strength and state density on the linear and non-linear spectra of our sample.
	
%\section{Conclusions}

In summary, we performed TSF measurements to explore the electronic structure of MoS$_2$ thin films.
Our TSF measurements uncover a conspicuous difference between absorption, SHG, and TSF spectra: the C feature is prominent in absorption and SHG but not TSF.
We address this conundrum by extracting the spectrally dependent dipole and JDOS using all three spectra.  
We find the differences in the spectra arise because the C feature has a large JDOS and small dipole compared to the A and B features.
We hope our measurements and analysis catalyze a renewed interest in elucidating the full spectral features of semiconductors by combining the results of many orders of complementary spectroscopies.
Our measurements demonstrate the utility of non-linear, sum-frequency spectroscopies of semiconductor nanostructures over a wide range of excitation frequencies. 
In the future, our work can be extended and to examine the coupling of multiple transitions which originate at the same point in the BZ and thus elucidate how different conduction or valence bands interact with each other.

\section*{Supplementary Information}

See Supplemental Information for synthesis and characterization of our MoS\textsubscript{2} thin film, more discussion of our ultrafast instrument and its calibration, discussion of data normalization scheme, discussion of our model, and additional simulation details

\begin{acknowledgements}
	This work was supported by the Department of Energy, Office of Basic Energy Sciences, Division of Materials Sciences and Engineering, under award DE--FG02--09ER46664.	
	The authors thank K. Lloyd for performing AFM measurements and T. Pedersen for sharing the SHG data used in this work. 
\end{acknowledgements}

\bibliography{mybib}

\begin{thebibliography}{63}%
\makeatletter
\providecommand \@ifxundefined [1]{%
 \@ifx{#1\undefined}
}%
\providecommand \@ifnum [1]{%
 \ifnum #1\expandafter \@firstoftwo
 \else \expandafter \@secondoftwo
 \fi
}%
\providecommand \@ifx [1]{%
 \ifx #1\expandafter \@firstoftwo
 \else \expandafter \@secondoftwo
 \fi
}%
\providecommand \natexlab [1]{#1}%
\providecommand \enquote  [1]{``#1''}%
\providecommand \bibnamefont  [1]{#1}%
\providecommand \bibfnamefont [1]{#1}%
\providecommand \citenamefont [1]{#1}%
\providecommand \href@noop [0]{\@secondoftwo}%
\providecommand \href [0]{\begingroup \@sanitize@url \@href}%
\providecommand \@href[1]{\@@startlink{#1}\@@href}%
\providecommand \@@href[1]{\endgroup#1\@@endlink}%
\providecommand \@sanitize@url [0]{\catcode `\\12\catcode `\$12\catcode
  `\&12\catcode `\#12\catcode `\^12\catcode `\_12\catcode `\%12\relax}%
\providecommand \@@startlink[1]{}%
\providecommand \@@endlink[0]{}%
\providecommand \url  [0]{\begingroup\@sanitize@url \@url }%
\providecommand \@url [1]{\endgroup\@href {#1}{\urlprefix }}%
\providecommand \urlprefix  [0]{URL }%
\providecommand \Eprint [0]{\href }%
\providecommand \doibase [0]{http://dx.doi.org/}%
\providecommand \selectlanguage [0]{\@gobble}%
\providecommand \bibinfo  [0]{\@secondoftwo}%
\providecommand \bibfield  [0]{\@secondoftwo}%
\providecommand \translation [1]{[#1]}%
\providecommand \BibitemOpen [0]{}%
\providecommand \bibitemStop [0]{}%
\providecommand \bibitemNoStop [0]{.\EOS\space}%
\providecommand \EOS [0]{\spacefactor3000\relax}%
\providecommand \BibitemShut  [1]{\csname bibitem#1\endcsname}%
\let\auto@bib@innerbib\@empty
%</preamble>
\bibitem [{\citenamefont {Cundiff}(2008)}]{Cundiff_2008}%
  \BibitemOpen
  \bibfield  {author} {\bibinfo {author} {\bibfnamefont {S.~T.}\ \bibnamefont
  {Cundiff}},\ }\href {\doibase 10.1364/oe.16.004639} {\bibfield  {journal}
  {\bibinfo  {journal} {Optics Express}\ }\textbf {\bibinfo {volume} {16}},\
  \bibinfo {pages} {4639} (\bibinfo {year} {2008})}\BibitemShut {NoStop}%
\bibitem [{\citenamefont {Moody}\ and\ \citenamefont
  {Cundiff}(2017)}]{Moody_Cundiff_2017}%
  \BibitemOpen
  \bibfield  {author} {\bibinfo {author} {\bibfnamefont {G.}~\bibnamefont
  {Moody}}\ and\ \bibinfo {author} {\bibfnamefont {S.~T.}\ \bibnamefont
  {Cundiff}},\ }\href {\doibase 10.1080/23746149.2017.1346482} {\bibfield
  {journal} {\bibinfo  {journal} {Advances in Physics: X}\ }\textbf {\bibinfo
  {volume} {2}},\ \bibinfo {pages} {641} (\bibinfo {year} {2017})}\BibitemShut
  {NoStop}%
\bibitem [{\citenamefont {Neff-Mallon}\ and\ \citenamefont
  {Wright}(2017)}]{NeffMallon_Wright_2017}%
  \BibitemOpen
  \bibfield  {author} {\bibinfo {author} {\bibfnamefont {N.~A.}\ \bibnamefont
  {Neff-Mallon}}\ and\ \bibinfo {author} {\bibfnamefont {J.~C.}\ \bibnamefont
  {Wright}},\ }\href {\doibase 10.1021/acs.analchem.7b02917} {\bibfield
  {journal} {\bibinfo  {journal} {Analytical Chemistry}\ }\textbf {\bibinfo
  {volume} {89}},\ \bibinfo {pages} {13182} (\bibinfo {year}
  {2017})}\BibitemShut {NoStop}%
\bibitem [{\citenamefont {Boyle}, \citenamefont {Pakoulev},\ and\ \citenamefont
  {Wright}(2013)}]{Boyle_Wright_2013}%
  \BibitemOpen
  \bibfield  {author} {\bibinfo {author} {\bibfnamefont {E.~S.}\ \bibnamefont
  {Boyle}}, \bibinfo {author} {\bibfnamefont {A.~V.}\ \bibnamefont {Pakoulev}},
  \ and\ \bibinfo {author} {\bibfnamefont {J.~C.}\ \bibnamefont {Wright}},\
  }\href {\doibase 10.1021/jp404713x} {\bibfield  {journal} {\bibinfo
  {journal} {The Journal of Physical Chemistry A}\ }\textbf {\bibinfo {volume}
  {117}},\ \bibinfo {pages} {5578} (\bibinfo {year} {2013})}\BibitemShut
  {NoStop}%
\bibitem [{\citenamefont {Boyle}, \citenamefont {Neff-Mallon},\ and\
  \citenamefont {Wright}(2013)}]{Boyle_Wright_2013_001}%
  \BibitemOpen
  \bibfield  {author} {\bibinfo {author} {\bibfnamefont {E.~S.}\ \bibnamefont
  {Boyle}}, \bibinfo {author} {\bibfnamefont {N.~A.}\ \bibnamefont
  {Neff-Mallon}}, \ and\ \bibinfo {author} {\bibfnamefont {J.~C.}\ \bibnamefont
  {Wright}},\ }\href {\doibase 10.1021/jp409377a} {\bibfield  {journal}
  {\bibinfo  {journal} {The Journal of Physical Chemistry A}\ }\textbf
  {\bibinfo {volume} {117}},\ \bibinfo {pages} {12401} (\bibinfo {year}
  {2013})}\BibitemShut {NoStop}%
\bibitem [{\citenamefont {Boyle}\ \emph {et~al.}(2014)\citenamefont {Boyle},
  \citenamefont {Neff-Mallon}, \citenamefont {Handali},\ and\ \citenamefont
  {Wright}}]{Boyle_Wright_2014}%
  \BibitemOpen
  \bibfield  {author} {\bibinfo {author} {\bibfnamefont {E.~S.}\ \bibnamefont
  {Boyle}}, \bibinfo {author} {\bibfnamefont {N.~A.}\ \bibnamefont
  {Neff-Mallon}}, \bibinfo {author} {\bibfnamefont {J.~D.}\ \bibnamefont
  {Handali}}, \ and\ \bibinfo {author} {\bibfnamefont {J.~C.}\ \bibnamefont
  {Wright}},\ }\href {\doibase 10.1021/jp5018554} {\bibfield  {journal}
  {\bibinfo  {journal} {The Journal of Physical Chemistry A}\ }\textbf
  {\bibinfo {volume} {118}},\ \bibinfo {pages} {3112} (\bibinfo {year}
  {2014})}\BibitemShut {NoStop}%
\bibitem [{\citenamefont {Grechko}\ \emph {et~al.}(2018)\citenamefont
  {Grechko}, \citenamefont {Hasegawa}, \citenamefont {D'Angelo}, \citenamefont
  {Ito}, \citenamefont {Turchinovich}, \citenamefont {Nagata},\ and\
  \citenamefont {Bonn}}]{Grechko_Bonn_2018}%
  \BibitemOpen
  \bibfield  {author} {\bibinfo {author} {\bibfnamefont {M.}~\bibnamefont
  {Grechko}}, \bibinfo {author} {\bibfnamefont {T.}~\bibnamefont {Hasegawa}},
  \bibinfo {author} {\bibfnamefont {F.}~\bibnamefont {D'Angelo}}, \bibinfo
  {author} {\bibfnamefont {H.}~\bibnamefont {Ito}}, \bibinfo {author}
  {\bibfnamefont {D.}~\bibnamefont {Turchinovich}}, \bibinfo {author}
  {\bibfnamefont {Y.}~\bibnamefont {Nagata}}, \ and\ \bibinfo {author}
  {\bibfnamefont {M.}~\bibnamefont {Bonn}},\ }\href {\doibase
  10.1038/s41467-018-03303-y} {\bibfield  {journal} {\bibinfo  {journal}
  {Nature Communications}\ }\textbf {\bibinfo {volume} {9}} (\bibinfo {year}
  {2018}),\ 10.1038/s41467-018-03303-y}\BibitemShut {NoStop}%
\bibitem [{\citenamefont {Mak}\ \emph {et~al.}(2010)\citenamefont {Mak},
  \citenamefont {Lee}, \citenamefont {Hone}, \citenamefont {Shan},\ and\
  \citenamefont {Heinz}}]{Mak_Heinz_2010}%
  \BibitemOpen
  \bibfield  {author} {\bibinfo {author} {\bibfnamefont {K.~F.}\ \bibnamefont
  {Mak}}, \bibinfo {author} {\bibfnamefont {C.}~\bibnamefont {Lee}}, \bibinfo
  {author} {\bibfnamefont {J.}~\bibnamefont {Hone}}, \bibinfo {author}
  {\bibfnamefont {J.}~\bibnamefont {Shan}}, \ and\ \bibinfo {author}
  {\bibfnamefont {T.~F.}\ \bibnamefont {Heinz}},\ }\href {\doibase
  10.1103/physrevlett.105.136805} {\bibfield  {journal} {\bibinfo  {journal}
  {Physical Review Letters}\ }\textbf {\bibinfo {volume} {105}} (\bibinfo
  {year} {2010}),\ 10.1103/physrevlett.105.136805}\BibitemShut {NoStop}%
\bibitem [{\citenamefont {Ellis}, \citenamefont {Lucero},\ and\ \citenamefont
  {Scuseria}(2011)}]{Ellis_Scuseria_2011}%
  \BibitemOpen
  \bibfield  {author} {\bibinfo {author} {\bibfnamefont {J.~K.}\ \bibnamefont
  {Ellis}}, \bibinfo {author} {\bibfnamefont {M.~J.}\ \bibnamefont {Lucero}}, \
  and\ \bibinfo {author} {\bibfnamefont {G.~E.}\ \bibnamefont {Scuseria}},\
  }\href {\doibase 10.1063/1.3672219} {\bibfield  {journal} {\bibinfo
  {journal} {Applied Physics Letters}\ }\textbf {\bibinfo {volume} {99}},\
  \bibinfo {pages} {261908} (\bibinfo {year} {2011})}\BibitemShut {NoStop}%
\bibitem [{\citenamefont {Wang}\ \emph {et~al.}(2012)\citenamefont {Wang},
  \citenamefont {Kalantar-Zadeh}, \citenamefont {Kis}, \citenamefont
  {Coleman},\ and\ \citenamefont {Strano}}]{Wang_Strano_2012}%
  \BibitemOpen
  \bibfield  {author} {\bibinfo {author} {\bibfnamefont {Q.~H.}\ \bibnamefont
  {Wang}}, \bibinfo {author} {\bibfnamefont {K.}~\bibnamefont
  {Kalantar-Zadeh}}, \bibinfo {author} {\bibfnamefont {A.}~\bibnamefont {Kis}},
  \bibinfo {author} {\bibfnamefont {J.~N.}\ \bibnamefont {Coleman}}, \ and\
  \bibinfo {author} {\bibfnamefont {M.~S.}\ \bibnamefont {Strano}},\ }\href
  {\doibase 10.1038/nnano.2012.193} {\bibfield  {journal} {\bibinfo  {journal}
  {Nature Nanotechnology}\ }\textbf {\bibinfo {volume} {7}},\ \bibinfo {pages}
  {699} (\bibinfo {year} {2012})}\BibitemShut {NoStop}%
\bibitem [{\citenamefont {Xia}\ \emph {et~al.}(2014)\citenamefont {Xia},
  \citenamefont {Wang}, \citenamefont {Xiao}, \citenamefont {Dubey},\ and\
  \citenamefont {Ramasubramaniam}}]{Xia_Ramasubramaniam_2014}%
  \BibitemOpen
  \bibfield  {author} {\bibinfo {author} {\bibfnamefont {F.}~\bibnamefont
  {Xia}}, \bibinfo {author} {\bibfnamefont {H.}~\bibnamefont {Wang}}, \bibinfo
  {author} {\bibfnamefont {D.}~\bibnamefont {Xiao}}, \bibinfo {author}
  {\bibfnamefont {M.}~\bibnamefont {Dubey}}, \ and\ \bibinfo {author}
  {\bibfnamefont {A.}~\bibnamefont {Ramasubramaniam}},\ }\href {\doibase
  10.1038/nphoton.2014.271} {\bibfield  {journal} {\bibinfo  {journal} {Nature
  Photonics}\ }\textbf {\bibinfo {volume} {8}},\ \bibinfo {pages} {899}
  (\bibinfo {year} {2014})}\BibitemShut {NoStop}%
\bibitem [{\citenamefont {Mak}\ and\ \citenamefont
  {Shan}(2016)}]{Mak_Shan_2016}%
  \BibitemOpen
  \bibfield  {author} {\bibinfo {author} {\bibfnamefont {K.~F.}\ \bibnamefont
  {Mak}}\ and\ \bibinfo {author} {\bibfnamefont {J.}~\bibnamefont {Shan}},\
  }\href {\doibase 10.1038/nphoton.2015.282} {\bibfield  {journal} {\bibinfo
  {journal} {Nature Photonics}\ }\textbf {\bibinfo {volume} {10}},\ \bibinfo
  {pages} {216} (\bibinfo {year} {2016})}\BibitemShut {NoStop}%
\bibitem [{\citenamefont {Molina-S{\'{a}}nchez}\ \emph
  {et~al.}(2013)\citenamefont {Molina-S{\'{a}}nchez}, \citenamefont {Sangalli},
  \citenamefont {Hummer}, \citenamefont {Marini},\ and\ \citenamefont
  {Wirtz}}]{MolinaSanchez_Wirtz_2013}%
  \BibitemOpen
  \bibfield  {author} {\bibinfo {author} {\bibfnamefont {A.}~\bibnamefont
  {Molina-S{\'{a}}nchez}}, \bibinfo {author} {\bibfnamefont {D.}~\bibnamefont
  {Sangalli}}, \bibinfo {author} {\bibfnamefont {K.}~\bibnamefont {Hummer}},
  \bibinfo {author} {\bibfnamefont {A.}~\bibnamefont {Marini}}, \ and\ \bibinfo
  {author} {\bibfnamefont {L.}~\bibnamefont {Wirtz}},\ }\href {\doibase
  10.1103/physrevb.88.045412} {\bibfield  {journal} {\bibinfo  {journal}
  {Physical Review B}\ }\textbf {\bibinfo {volume} {88}} (\bibinfo {year}
  {2013}),\ 10.1103/physrevb.88.045412}\BibitemShut {NoStop}%
\bibitem [{\citenamefont {Li}\ \emph {et~al.}(2014)\citenamefont {Li},
  \citenamefont {Chernikov}, \citenamefont {Zhang}, \citenamefont {Rigosi},
  \citenamefont {Hill}, \citenamefont {van~der Zande}, \citenamefont {Chenet},
  \citenamefont {Shih}, \citenamefont {Hone},\ and\ \citenamefont
  {Heinz}}]{Li_Heinz_2014}%
  \BibitemOpen
  \bibfield  {author} {\bibinfo {author} {\bibfnamefont {Y.}~\bibnamefont
  {Li}}, \bibinfo {author} {\bibfnamefont {A.}~\bibnamefont {Chernikov}},
  \bibinfo {author} {\bibfnamefont {X.}~\bibnamefont {Zhang}}, \bibinfo
  {author} {\bibfnamefont {A.}~\bibnamefont {Rigosi}}, \bibinfo {author}
  {\bibfnamefont {H.~M.}\ \bibnamefont {Hill}}, \bibinfo {author}
  {\bibfnamefont {A.~M.}\ \bibnamefont {van~der Zande}}, \bibinfo {author}
  {\bibfnamefont {D.~A.}\ \bibnamefont {Chenet}}, \bibinfo {author}
  {\bibfnamefont {E.-M.}\ \bibnamefont {Shih}}, \bibinfo {author}
  {\bibfnamefont {J.}~\bibnamefont {Hone}}, \ and\ \bibinfo {author}
  {\bibfnamefont {T.~F.}\ \bibnamefont {Heinz}},\ }\href {\doibase
  10.1103/physrevb.90.205422} {\bibfield  {journal} {\bibinfo  {journal}
  {Physical Review B}\ }\textbf {\bibinfo {volume} {90}} (\bibinfo {year}
  {2014}),\ 10.1103/physrevb.90.205422}\BibitemShut {NoStop}%
\bibitem [{\citenamefont {Qiu}, \citenamefont {da~Jornada},\ and\ \citenamefont
  {Louie}(2013)}]{Qiu_Louie_2013}%
  \BibitemOpen
  \bibfield  {author} {\bibinfo {author} {\bibfnamefont {D.~Y.}\ \bibnamefont
  {Qiu}}, \bibinfo {author} {\bibfnamefont {F.~H.}\ \bibnamefont {da~Jornada}},
  \ and\ \bibinfo {author} {\bibfnamefont {S.~G.}\ \bibnamefont {Louie}},\
  }\href {\doibase 10.1103/physrevlett.111.216805} {\bibfield  {journal}
  {\bibinfo  {journal} {Physical Review Letters}\ }\textbf {\bibinfo {volume}
  {111}} (\bibinfo {year} {2013}),\ 10.1103/physrevlett.111.216805}\BibitemShut
  {NoStop}%
\bibitem [{\citenamefont {He}\ \emph {et~al.}(2014)\citenamefont {He},
  \citenamefont {Kumar}, \citenamefont {Zhao}, \citenamefont {Wang},
  \citenamefont {Mak}, \citenamefont {Zhao},\ and\ \citenamefont
  {Shan}}]{He_Shan_2014}%
  \BibitemOpen
  \bibfield  {author} {\bibinfo {author} {\bibfnamefont {K.}~\bibnamefont
  {He}}, \bibinfo {author} {\bibfnamefont {N.}~\bibnamefont {Kumar}}, \bibinfo
  {author} {\bibfnamefont {L.}~\bibnamefont {Zhao}}, \bibinfo {author}
  {\bibfnamefont {Z.}~\bibnamefont {Wang}}, \bibinfo {author} {\bibfnamefont
  {K.~F.}\ \bibnamefont {Mak}}, \bibinfo {author} {\bibfnamefont
  {H.}~\bibnamefont {Zhao}}, \ and\ \bibinfo {author} {\bibfnamefont
  {J.}~\bibnamefont {Shan}},\ }\href {\doibase 10.1103/physrevlett.113.026803}
  {\bibfield  {journal} {\bibinfo  {journal} {Physical Review Letters}\
  }\textbf {\bibinfo {volume} {113}} (\bibinfo {year} {2014}),\
  10.1103/physrevlett.113.026803}\BibitemShut {NoStop}%
\bibitem [{\citenamefont {Saigal}, \citenamefont {Sugunakar},\ and\
  \citenamefont {Ghosh}(2016)}]{Saigal_Ghosh_2016}%
  \BibitemOpen
  \bibfield  {author} {\bibinfo {author} {\bibfnamefont {N.}~\bibnamefont
  {Saigal}}, \bibinfo {author} {\bibfnamefont {V.}~\bibnamefont {Sugunakar}}, \
  and\ \bibinfo {author} {\bibfnamefont {S.}~\bibnamefont {Ghosh}},\ }\href
  {\doibase 10.1063/1.4945047} {\bibfield  {journal} {\bibinfo  {journal}
  {Applied Physics Letters}\ }\textbf {\bibinfo {volume} {108}},\ \bibinfo
  {pages} {132105} (\bibinfo {year} {2016})}\BibitemShut {NoStop}%
\bibitem [{\citenamefont {Kopaczek}\ \emph {et~al.}(2016)\citenamefont
  {Kopaczek}, \citenamefont {Polak}, \citenamefont {Scharoch}, \citenamefont
  {Wu}, \citenamefont {Chen}, \citenamefont {Tongay},\ and\ \citenamefont
  {Kudrawiec}}]{Kopaczek_Kudrawiec_2016}%
  \BibitemOpen
  \bibfield  {author} {\bibinfo {author} {\bibfnamefont {J.}~\bibnamefont
  {Kopaczek}}, \bibinfo {author} {\bibfnamefont {M.~P.}\ \bibnamefont {Polak}},
  \bibinfo {author} {\bibfnamefont {P.}~\bibnamefont {Scharoch}}, \bibinfo
  {author} {\bibfnamefont {K.}~\bibnamefont {Wu}}, \bibinfo {author}
  {\bibfnamefont {B.}~\bibnamefont {Chen}}, \bibinfo {author} {\bibfnamefont
  {S.}~\bibnamefont {Tongay}}, \ and\ \bibinfo {author} {\bibfnamefont
  {R.}~\bibnamefont {Kudrawiec}},\ }\href {\doibase 10.1063/1.4954157}
  {\bibfield  {journal} {\bibinfo  {journal} {Journal of Applied Physics}\
  }\textbf {\bibinfo {volume} {119}},\ \bibinfo {pages} {235705} (\bibinfo
  {year} {2016})}\BibitemShut {NoStop}%
\bibitem [{\citenamefont {Britnell}\ \emph {et~al.}(2013)\citenamefont
  {Britnell}, \citenamefont {Ribeiro}, \citenamefont {Eckmann}, \citenamefont
  {Jalil}, \citenamefont {Belle}, \citenamefont {Mishchenko}, \citenamefont
  {Kim}, \citenamefont {Gorbachev}, \citenamefont {Georgiou}, \citenamefont
  {Morozov}, \citenamefont {Grigorenko}, \citenamefont {Geim}, \citenamefont
  {Casiraghi}, \citenamefont {Neto},\ and\ \citenamefont
  {Novoselov}}]{Britnell_Novoselov_2013}%
  \BibitemOpen
  \bibfield  {author} {\bibinfo {author} {\bibfnamefont {L.}~\bibnamefont
  {Britnell}}, \bibinfo {author} {\bibfnamefont {R.~M.}\ \bibnamefont
  {Ribeiro}}, \bibinfo {author} {\bibfnamefont {A.}~\bibnamefont {Eckmann}},
  \bibinfo {author} {\bibfnamefont {R.}~\bibnamefont {Jalil}}, \bibinfo
  {author} {\bibfnamefont {B.~D.}\ \bibnamefont {Belle}}, \bibinfo {author}
  {\bibfnamefont {A.}~\bibnamefont {Mishchenko}}, \bibinfo {author}
  {\bibfnamefont {Y.-J.}\ \bibnamefont {Kim}}, \bibinfo {author} {\bibfnamefont
  {R.~V.}\ \bibnamefont {Gorbachev}}, \bibinfo {author} {\bibfnamefont
  {T.}~\bibnamefont {Georgiou}}, \bibinfo {author} {\bibfnamefont {S.~V.}\
  \bibnamefont {Morozov}}, \bibinfo {author} {\bibfnamefont {A.~N.}\
  \bibnamefont {Grigorenko}}, \bibinfo {author} {\bibfnamefont {A.~K.}\
  \bibnamefont {Geim}}, \bibinfo {author} {\bibfnamefont {C.}~\bibnamefont
  {Casiraghi}}, \bibinfo {author} {\bibfnamefont {A.~H.~C.}\ \bibnamefont
  {Neto}}, \ and\ \bibinfo {author} {\bibfnamefont {K.~S.}\ \bibnamefont
  {Novoselov}},\ }\href {\doibase 10.1126/science.1235547} {\bibfield
  {journal} {\bibinfo  {journal} {Science}\ }\textbf {\bibinfo {volume}
  {340}},\ \bibinfo {pages} {1311} (\bibinfo {year} {2013})}\BibitemShut
  {NoStop}%
\bibitem [{\citenamefont {Carvalho}, \citenamefont {Ribeiro},\ and\
  \citenamefont {Neto}(2013)}]{Carvalho_Neto_2013}%
  \BibitemOpen
  \bibfield  {author} {\bibinfo {author} {\bibfnamefont {A.}~\bibnamefont
  {Carvalho}}, \bibinfo {author} {\bibfnamefont {R.~M.}\ \bibnamefont
  {Ribeiro}}, \ and\ \bibinfo {author} {\bibfnamefont {A.~H.~C.}\ \bibnamefont
  {Neto}},\ }\href {\doibase 10.1103/physrevb.88.115205} {\bibfield  {journal}
  {\bibinfo  {journal} {Physical Review B}\ }\textbf {\bibinfo {volume} {88}}
  (\bibinfo {year} {2013}),\ 10.1103/physrevb.88.115205}\BibitemShut {NoStop}%
\bibitem [{\citenamefont {Jeong}\ \emph {et~al.}(2018)\citenamefont {Jeong},
  \citenamefont {Choi}, \citenamefont {Jeong}, \citenamefont {Park},
  \citenamefont {Kim},\ and\ \citenamefont {Cho}}]{Jeong_Cho_2018}%
  \BibitemOpen
  \bibfield  {author} {\bibinfo {author} {\bibfnamefont {J.}~\bibnamefont
  {Jeong}}, \bibinfo {author} {\bibfnamefont {Y.-H.}\ \bibnamefont {Choi}},
  \bibinfo {author} {\bibfnamefont {K.}~\bibnamefont {Jeong}}, \bibinfo
  {author} {\bibfnamefont {H.}~\bibnamefont {Park}}, \bibinfo {author}
  {\bibfnamefont {D.}~\bibnamefont {Kim}}, \ and\ \bibinfo {author}
  {\bibfnamefont {M.-H.}\ \bibnamefont {Cho}},\ }\href {\doibase
  10.1103/physrevb.97.075433} {\bibfield  {journal} {\bibinfo  {journal}
  {Physical Review B}\ }\textbf {\bibinfo {volume} {97}} (\bibinfo {year}
  {2018}),\ 10.1103/physrevb.97.075433}\BibitemShut {NoStop}%
\bibitem [{\citenamefont {Bieniek}\ \emph {et~al.}(2018)\citenamefont
  {Bieniek}, \citenamefont {Korkusi{\'{n}}ski}, \citenamefont {Szulakowska},
  \citenamefont {Potasz}, \citenamefont {Ozfidan},\ and\ \citenamefont
  {Hawrylak}}]{Bieniek_Hawrylak_2018}%
  \BibitemOpen
  \bibfield  {author} {\bibinfo {author} {\bibfnamefont {M.}~\bibnamefont
  {Bieniek}}, \bibinfo {author} {\bibfnamefont {M.}~\bibnamefont
  {Korkusi{\'{n}}ski}}, \bibinfo {author} {\bibfnamefont {L.}~\bibnamefont
  {Szulakowska}}, \bibinfo {author} {\bibfnamefont {P.}~\bibnamefont {Potasz}},
  \bibinfo {author} {\bibfnamefont {I.}~\bibnamefont {Ozfidan}}, \ and\
  \bibinfo {author} {\bibfnamefont {P.}~\bibnamefont {Hawrylak}},\ }\href
  {\doibase 10.1103/physrevb.97.085153} {\bibfield  {journal} {\bibinfo
  {journal} {Physical Review B}\ }\textbf {\bibinfo {volume} {97}} (\bibinfo
  {year} {2018}),\ 10.1103/physrevb.97.085153}\BibitemShut {NoStop}%
\bibitem [{\citenamefont {Boyd}(2008)}]{Boyd_2008}%
  \BibitemOpen
  \bibfield  {author} {\bibinfo {author} {\bibfnamefont {R.~W.}\ \bibnamefont
  {Boyd}},\ }\href@noop {} {\emph {\bibinfo {title} {Nonlinear Optics}}},\
  \bibinfo {edition} {3rd}\ ed.\ (\bibinfo  {publisher} {Academic Press},\
  \bibinfo {year} {2008})\BibitemShut {NoStop}%
\bibitem [{\citenamefont {Bloembergen}\ and\ \citenamefont
  {Shen}(1964)}]{Bloembergen_Shen_1964}%
  \BibitemOpen
  \bibfield  {author} {\bibinfo {author} {\bibfnamefont {N.}~\bibnamefont
  {Bloembergen}}\ and\ \bibinfo {author} {\bibfnamefont {Y.~R.}\ \bibnamefont
  {Shen}},\ }\href {\doibase 10.1103/physrev.133.a37} {\bibfield  {journal}
  {\bibinfo  {journal} {Physical Review}\ }\textbf {\bibinfo {volume} {133}},\
  \bibinfo {pages} {A37} (\bibinfo {year} {1964})}\BibitemShut {NoStop}%
\bibitem [{\citenamefont {Li}\ \emph {et~al.}(2013)\citenamefont {Li},
  \citenamefont {Rao}, \citenamefont {Mak}, \citenamefont {You}, \citenamefont
  {Wang}, \citenamefont {Dean},\ and\ \citenamefont {Heinz}}]{Li_Heinz_2013}%
  \BibitemOpen
  \bibfield  {author} {\bibinfo {author} {\bibfnamefont {Y.}~\bibnamefont
  {Li}}, \bibinfo {author} {\bibfnamefont {Y.}~\bibnamefont {Rao}}, \bibinfo
  {author} {\bibfnamefont {K.~F.}\ \bibnamefont {Mak}}, \bibinfo {author}
  {\bibfnamefont {Y.}~\bibnamefont {You}}, \bibinfo {author} {\bibfnamefont
  {S.}~\bibnamefont {Wang}}, \bibinfo {author} {\bibfnamefont {C.~R.}\
  \bibnamefont {Dean}}, \ and\ \bibinfo {author} {\bibfnamefont {T.~F.}\
  \bibnamefont {Heinz}},\ }\href {\doibase 10.1021/nl401561r} {\bibfield
  {journal} {\bibinfo  {journal} {Nano Letters}\ }\textbf {\bibinfo {volume}
  {13}},\ \bibinfo {pages} {3329} (\bibinfo {year} {2013})}\BibitemShut
  {NoStop}%
\bibitem [{\citenamefont {Malard}\ \emph {et~al.}(2013)\citenamefont {Malard},
  \citenamefont {Alencar}, \citenamefont {Barboza}, \citenamefont {Mak},\ and\
  \citenamefont {de~Paula}}]{Malard_dePaula_2013}%
  \BibitemOpen
  \bibfield  {author} {\bibinfo {author} {\bibfnamefont {L.~M.}\ \bibnamefont
  {Malard}}, \bibinfo {author} {\bibfnamefont {T.~V.}\ \bibnamefont {Alencar}},
  \bibinfo {author} {\bibfnamefont {A.~P.~M.}\ \bibnamefont {Barboza}},
  \bibinfo {author} {\bibfnamefont {K.~F.}\ \bibnamefont {Mak}}, \ and\
  \bibinfo {author} {\bibfnamefont {A.~M.}\ \bibnamefont {de~Paula}},\ }\href
  {\doibase 10.1103/physrevb.87.201401} {\bibfield  {journal} {\bibinfo
  {journal} {Physical Review B}\ }\textbf {\bibinfo {volume} {87}} (\bibinfo
  {year} {2013}),\ 10.1103/physrevb.87.201401}\BibitemShut {NoStop}%
\bibitem [{\citenamefont {Kumar}\ \emph {et~al.}(2013)\citenamefont {Kumar},
  \citenamefont {Najmaei}, \citenamefont {Cui}, \citenamefont {Ceballos},
  \citenamefont {Ajayan}, \citenamefont {Lou},\ and\ \citenamefont
  {Zhao}}]{Kumar_Zhao_2013}%
  \BibitemOpen
  \bibfield  {author} {\bibinfo {author} {\bibfnamefont {N.}~\bibnamefont
  {Kumar}}, \bibinfo {author} {\bibfnamefont {S.}~\bibnamefont {Najmaei}},
  \bibinfo {author} {\bibfnamefont {Q.}~\bibnamefont {Cui}}, \bibinfo {author}
  {\bibfnamefont {F.}~\bibnamefont {Ceballos}}, \bibinfo {author}
  {\bibfnamefont {P.~M.}\ \bibnamefont {Ajayan}}, \bibinfo {author}
  {\bibfnamefont {J.}~\bibnamefont {Lou}}, \ and\ \bibinfo {author}
  {\bibfnamefont {H.}~\bibnamefont {Zhao}},\ }\href {\doibase
  10.1103/physrevb.87.161403} {\bibfield  {journal} {\bibinfo  {journal}
  {Physical Review B}\ }\textbf {\bibinfo {volume} {87}} (\bibinfo {year}
  {2013}),\ 10.1103/physrevb.87.161403}\BibitemShut {NoStop}%
\bibitem [{\citenamefont {Wang}\ \emph {et~al.}(2013)\citenamefont {Wang},
  \citenamefont {Chien}, \citenamefont {Kumar}, \citenamefont {Kumar},
  \citenamefont {Chiu},\ and\ \citenamefont {Zhao}}]{Wang_Zhao_2013}%
  \BibitemOpen
  \bibfield  {author} {\bibinfo {author} {\bibfnamefont {R.}~\bibnamefont
  {Wang}}, \bibinfo {author} {\bibfnamefont {H.-C.}\ \bibnamefont {Chien}},
  \bibinfo {author} {\bibfnamefont {J.}~\bibnamefont {Kumar}}, \bibinfo
  {author} {\bibfnamefont {N.}~\bibnamefont {Kumar}}, \bibinfo {author}
  {\bibfnamefont {H.-Y.}\ \bibnamefont {Chiu}}, \ and\ \bibinfo {author}
  {\bibfnamefont {H.}~\bibnamefont {Zhao}},\ }\href {\doibase
  10.1021/am4042542} {\bibfield  {journal} {\bibinfo  {journal} {{ACS} Applied
  Materials {\&} Interfaces}\ }\textbf {\bibinfo {volume} {6}},\ \bibinfo
  {pages} {314} (\bibinfo {year} {2013})}\BibitemShut {NoStop}%
\bibitem [{\citenamefont {Trolle}, \citenamefont {Seifert},\ and\ \citenamefont
  {Pedersen}(2014)}]{Trolle_Pedersen_2014}%
  \BibitemOpen
  \bibfield  {author} {\bibinfo {author} {\bibfnamefont {M.~L.}\ \bibnamefont
  {Trolle}}, \bibinfo {author} {\bibfnamefont {G.}~\bibnamefont {Seifert}}, \
  and\ \bibinfo {author} {\bibfnamefont {T.~G.}\ \bibnamefont {Pedersen}},\
  }\href {\doibase 10.1103/physrevb.89.235410} {\bibfield  {journal} {\bibinfo
  {journal} {Physical Review B}\ }\textbf {\bibinfo {volume} {89}} (\bibinfo
  {year} {2014}),\ 10.1103/physrevb.89.235410}\BibitemShut {NoStop}%
\bibitem [{\citenamefont {Gr\"{u}ning}\ and\ \citenamefont
  {Attaccalite}(2014)}]{Gruning_Attaccalite_2014}%
  \BibitemOpen
  \bibfield  {author} {\bibinfo {author} {\bibfnamefont {M.}~\bibnamefont
  {Gr\"{u}ning}}\ and\ \bibinfo {author} {\bibfnamefont {C.}~\bibnamefont
  {Attaccalite}},\ }\href {\doibase 10.1103/physrevb.89.081102} {\bibfield
  {journal} {\bibinfo  {journal} {Physical Review B}\ }\textbf {\bibinfo
  {volume} {89}} (\bibinfo {year} {2014}),\
  10.1103/physrevb.89.081102}\BibitemShut {NoStop}%
\bibitem [{\citenamefont {Clark}\ \emph {et~al.}(2014)\citenamefont {Clark},
  \citenamefont {Senthilkumar}, \citenamefont {Le}, \citenamefont {Weerawarne},
  \citenamefont {Shim}, \citenamefont {Jang}, \citenamefont {Shim},
  \citenamefont {Cho}, \citenamefont {Sim}, \citenamefont {Seong},
  \citenamefont {Rhim}, \citenamefont {Freeman}, \citenamefont {Chung},\ and\
  \citenamefont {Kim}}]{Clark_Kim_2014}%
  \BibitemOpen
  \bibfield  {author} {\bibinfo {author} {\bibfnamefont {D.~J.}\ \bibnamefont
  {Clark}}, \bibinfo {author} {\bibfnamefont {V.}~\bibnamefont {Senthilkumar}},
  \bibinfo {author} {\bibfnamefont {C.~T.}\ \bibnamefont {Le}}, \bibinfo
  {author} {\bibfnamefont {D.~L.}\ \bibnamefont {Weerawarne}}, \bibinfo
  {author} {\bibfnamefont {B.}~\bibnamefont {Shim}}, \bibinfo {author}
  {\bibfnamefont {J.~I.}\ \bibnamefont {Jang}}, \bibinfo {author}
  {\bibfnamefont {J.~H.}\ \bibnamefont {Shim}}, \bibinfo {author}
  {\bibfnamefont {J.}~\bibnamefont {Cho}}, \bibinfo {author} {\bibfnamefont
  {Y.}~\bibnamefont {Sim}}, \bibinfo {author} {\bibfnamefont {M.-J.}\
  \bibnamefont {Seong}}, \bibinfo {author} {\bibfnamefont {S.~H.}\ \bibnamefont
  {Rhim}}, \bibinfo {author} {\bibfnamefont {A.~J.}\ \bibnamefont {Freeman}},
  \bibinfo {author} {\bibfnamefont {K.-H.}\ \bibnamefont {Chung}}, \ and\
  \bibinfo {author} {\bibfnamefont {Y.~S.}\ \bibnamefont {Kim}},\ }\href
  {\doibase 10.1103/physrevb.90.121409} {\bibfield  {journal} {\bibinfo
  {journal} {Physical Review B}\ }\textbf {\bibinfo {volume} {90}} (\bibinfo
  {year} {2014}),\ 10.1103/physrevb.90.121409}\BibitemShut {NoStop}%
\bibitem [{\citenamefont {Trolle}\ \emph {et~al.}(2015)\citenamefont {Trolle},
  \citenamefont {Tsao}, \citenamefont {Pedersen},\ and\ \citenamefont
  {Pedersen}}]{Trolle_Pedersen_2015}%
  \BibitemOpen
  \bibfield  {author} {\bibinfo {author} {\bibfnamefont {M.~L.}\ \bibnamefont
  {Trolle}}, \bibinfo {author} {\bibfnamefont {Y.-C.}\ \bibnamefont {Tsao}},
  \bibinfo {author} {\bibfnamefont {K.}~\bibnamefont {Pedersen}}, \ and\
  \bibinfo {author} {\bibfnamefont {T.~G.}\ \bibnamefont {Pedersen}},\ }\href
  {\doibase 10.1103/physrevb.92.161409} {\bibfield  {journal} {\bibinfo
  {journal} {Physical Review B}\ }\textbf {\bibinfo {volume} {92}} (\bibinfo
  {year} {2015}),\ 10.1103/physrevb.92.161409}\BibitemShut {NoStop}%
\bibitem [{\citenamefont {Wang}\ \emph {et~al.}(2015)\citenamefont {Wang},
  \citenamefont {Marie}, \citenamefont {Gerber}, \citenamefont {Amand},
  \citenamefont {Lagarde}, \citenamefont {Bouet}, \citenamefont {Vidal},
  \citenamefont {Balocchi},\ and\ \citenamefont
  {Urbaszek}}]{Wang_Urbaszek_2015}%
  \BibitemOpen
  \bibfield  {author} {\bibinfo {author} {\bibfnamefont {G.}~\bibnamefont
  {Wang}}, \bibinfo {author} {\bibfnamefont {X.}~\bibnamefont {Marie}},
  \bibinfo {author} {\bibfnamefont {I.}~\bibnamefont {Gerber}}, \bibinfo
  {author} {\bibfnamefont {T.}~\bibnamefont {Amand}}, \bibinfo {author}
  {\bibfnamefont {D.}~\bibnamefont {Lagarde}}, \bibinfo {author} {\bibfnamefont
  {L.}~\bibnamefont {Bouet}}, \bibinfo {author} {\bibfnamefont
  {M.}~\bibnamefont {Vidal}}, \bibinfo {author} {\bibfnamefont
  {A.}~\bibnamefont {Balocchi}}, \ and\ \bibinfo {author} {\bibfnamefont
  {B.}~\bibnamefont {Urbaszek}},\ }\href {\doibase
  10.1103/physrevlett.114.097403} {\bibfield  {journal} {\bibinfo  {journal}
  {Physical Review Letters}\ }\textbf {\bibinfo {volume} {114}} (\bibinfo
  {year} {2015}),\ 10.1103/physrevlett.114.097403}\BibitemShut {NoStop}%
\bibitem [{\citenamefont {Sun}\ \emph {et~al.}(2016)\citenamefont {Sun},
  \citenamefont {Gu}, \citenamefont {Lei}, \citenamefont {Lau}, \citenamefont
  {Wong}, \citenamefont {Wong},\ and\ \citenamefont {Chan}}]{Sun_Chan_2016}%
  \BibitemOpen
  \bibfield  {author} {\bibinfo {author} {\bibfnamefont {J.}~\bibnamefont
  {Sun}}, \bibinfo {author} {\bibfnamefont {Y.-J.}\ \bibnamefont {Gu}},
  \bibinfo {author} {\bibfnamefont {D.~Y.}\ \bibnamefont {Lei}}, \bibinfo
  {author} {\bibfnamefont {S.~P.}\ \bibnamefont {Lau}}, \bibinfo {author}
  {\bibfnamefont {W.-T.}\ \bibnamefont {Wong}}, \bibinfo {author}
  {\bibfnamefont {K.-Y.}\ \bibnamefont {Wong}}, \ and\ \bibinfo {author}
  {\bibfnamefont {H.~L.-W.}\ \bibnamefont {Chan}},\ }\href {\doibase
  10.1021/acsphotonics.6b00682} {\bibfield  {journal} {\bibinfo  {journal}
  {{ACS} Photonics}\ }\textbf {\bibinfo {volume} {3}},\ \bibinfo {pages} {2434}
  (\bibinfo {year} {2016})}\BibitemShut {NoStop}%
\bibitem [{\citenamefont {Karvonen}\ \emph {et~al.}(2017)\citenamefont
  {Karvonen}, \citenamefont {S\"{a}yn\"{a}tjoki}, \citenamefont {Huttunen},
  \citenamefont {Autere}, \citenamefont {Amirsolaimani}, \citenamefont {Li},
  \citenamefont {Norwood}, \citenamefont {Peyghambarian}, \citenamefont
  {Lipsanen}, \citenamefont {Eda}, \citenamefont {Kieu},\ and\ \citenamefont
  {Sun}}]{Karvonen_Sun_2017}%
  \BibitemOpen
  \bibfield  {author} {\bibinfo {author} {\bibfnamefont {L.}~\bibnamefont
  {Karvonen}}, \bibinfo {author} {\bibfnamefont {A.}~\bibnamefont
  {S\"{a}yn\"{a}tjoki}}, \bibinfo {author} {\bibfnamefont {M.~J.}\ \bibnamefont
  {Huttunen}}, \bibinfo {author} {\bibfnamefont {A.}~\bibnamefont {Autere}},
  \bibinfo {author} {\bibfnamefont {B.}~\bibnamefont {Amirsolaimani}}, \bibinfo
  {author} {\bibfnamefont {S.}~\bibnamefont {Li}}, \bibinfo {author}
  {\bibfnamefont {R.~A.}\ \bibnamefont {Norwood}}, \bibinfo {author}
  {\bibfnamefont {N.}~\bibnamefont {Peyghambarian}}, \bibinfo {author}
  {\bibfnamefont {H.}~\bibnamefont {Lipsanen}}, \bibinfo {author}
  {\bibfnamefont {G.}~\bibnamefont {Eda}}, \bibinfo {author} {\bibfnamefont
  {K.}~\bibnamefont {Kieu}}, \ and\ \bibinfo {author} {\bibfnamefont
  {Z.}~\bibnamefont {Sun}},\ }\href {\doibase 10.1038/ncomms15714} {\bibfield
  {journal} {\bibinfo  {journal} {Nature Communications}\ }\textbf {\bibinfo
  {volume} {8}},\ \bibinfo {pages} {15714} (\bibinfo {year}
  {2017})}\BibitemShut {NoStop}%
\bibitem [{\citenamefont {Shearer}\ \emph {et~al.}(2017)\citenamefont
  {Shearer}, \citenamefont {Samad}, \citenamefont {Zhang}, \citenamefont
  {Zhao}, \citenamefont {Puretzky}, \citenamefont {Eliceiri}, \citenamefont
  {Wright}, \citenamefont {Hamers},\ and\ \citenamefont
  {Jin}}]{Shearer_Jin_2017}%
  \BibitemOpen
  \bibfield  {author} {\bibinfo {author} {\bibfnamefont {M.~J.}\ \bibnamefont
  {Shearer}}, \bibinfo {author} {\bibfnamefont {L.}~\bibnamefont {Samad}},
  \bibinfo {author} {\bibfnamefont {Y.}~\bibnamefont {Zhang}}, \bibinfo
  {author} {\bibfnamefont {Y.}~\bibnamefont {Zhao}}, \bibinfo {author}
  {\bibfnamefont {A.}~\bibnamefont {Puretzky}}, \bibinfo {author}
  {\bibfnamefont {K.~W.}\ \bibnamefont {Eliceiri}}, \bibinfo {author}
  {\bibfnamefont {J.~C.}\ \bibnamefont {Wright}}, \bibinfo {author}
  {\bibfnamefont {R.~J.}\ \bibnamefont {Hamers}}, \ and\ \bibinfo {author}
  {\bibfnamefont {S.}~\bibnamefont {Jin}},\ }\href {\doibase
  10.1021/jacs.6b12559} {\bibfield  {journal} {\bibinfo  {journal} {Journal of
  the American Chemical Society}\ }\textbf {\bibinfo {volume} {139}},\ \bibinfo
  {pages} {3496} (\bibinfo {year} {2017})}\BibitemShut {NoStop}%
\bibitem [{\citenamefont {Glazov}\ \emph {et~al.}(2017)\citenamefont {Glazov},
  \citenamefont {Golub}, \citenamefont {Wang}, \citenamefont {Marie},
  \citenamefont {Amand},\ and\ \citenamefont
  {Urbaszek}}]{Glazov_Urbaszek_2017}%
  \BibitemOpen
  \bibfield  {author} {\bibinfo {author} {\bibfnamefont {M.~M.}\ \bibnamefont
  {Glazov}}, \bibinfo {author} {\bibfnamefont {L.~E.}\ \bibnamefont {Golub}},
  \bibinfo {author} {\bibfnamefont {G.}~\bibnamefont {Wang}}, \bibinfo {author}
  {\bibfnamefont {X.}~\bibnamefont {Marie}}, \bibinfo {author} {\bibfnamefont
  {T.}~\bibnamefont {Amand}}, \ and\ \bibinfo {author} {\bibfnamefont
  {B.}~\bibnamefont {Urbaszek}},\ }\href {\doibase 10.1103/physrevb.95.035311}
  {\bibfield  {journal} {\bibinfo  {journal} {Physical Review B}\ }\textbf
  {\bibinfo {volume} {95}} (\bibinfo {year} {2017}),\
  10.1103/physrevb.95.035311}\BibitemShut {NoStop}%
\bibitem [{\citenamefont {Balla}\ \emph {et~al.}(2018)\citenamefont {Balla},
  \citenamefont {O'Brien}, \citenamefont {McEvoy}, \citenamefont {Duesberg},
  \citenamefont {Rigneault}, \citenamefont {Brasselet},\ and\ \citenamefont
  {McCloskey}}]{Balla_McCloskey_2018}%
  \BibitemOpen
  \bibfield  {author} {\bibinfo {author} {\bibfnamefont {N.~K.}\ \bibnamefont
  {Balla}}, \bibinfo {author} {\bibfnamefont {M.}~\bibnamefont {O'Brien}},
  \bibinfo {author} {\bibfnamefont {N.}~\bibnamefont {McEvoy}}, \bibinfo
  {author} {\bibfnamefont {G.~S.}\ \bibnamefont {Duesberg}}, \bibinfo {author}
  {\bibfnamefont {H.}~\bibnamefont {Rigneault}}, \bibinfo {author}
  {\bibfnamefont {S.}~\bibnamefont {Brasselet}}, \ and\ \bibinfo {author}
  {\bibfnamefont {D.}~\bibnamefont {McCloskey}},\ }\href {\doibase
  10.1021/acsphotonics.7b00912} {\bibfield  {journal} {\bibinfo  {journal}
  {{ACS} Photonics}\ } (\bibinfo {year} {2018}),\
  10.1021/acsphotonics.7b00912}\BibitemShut {NoStop}%
\bibitem [{\citenamefont {Wright}(2011)}]{Wright_2011}%
  \BibitemOpen
  \bibfield  {author} {\bibinfo {author} {\bibfnamefont {J.~C.}\ \bibnamefont
  {Wright}},\ }\href {\doibase 10.1146/annurev-physchem-032210-103551}
  {\bibfield  {journal} {\bibinfo  {journal} {Annual Review of Physical
  Chemistry}\ }\textbf {\bibinfo {volume} {62}},\ \bibinfo {pages} {209}
  (\bibinfo {year} {2011})}\BibitemShut {NoStop}%
\bibitem [{\citenamefont {Volkmer}, \citenamefont {Cheng},\ and\ \citenamefont
  {Xie}(2001)}]{Volkmer_Xie_2001}%
  \BibitemOpen
  \bibfield  {author} {\bibinfo {author} {\bibfnamefont {A.}~\bibnamefont
  {Volkmer}}, \bibinfo {author} {\bibfnamefont {J.-X.}\ \bibnamefont {Cheng}},
  \ and\ \bibinfo {author} {\bibfnamefont {X.~S.}\ \bibnamefont {Xie}},\ }\href
  {\doibase 10.1103/physrevlett.87.023901} {\bibfield  {journal} {\bibinfo
  {journal} {Physical Review Letters}\ }\textbf {\bibinfo {volume} {87}}
  (\bibinfo {year} {2001}),\ 10.1103/physrevlett.87.023901}\BibitemShut
  {NoStop}%
\bibitem [{\citenamefont {Czech}\ \emph {et~al.}(2015)\citenamefont {Czech},
  \citenamefont {Thompson}, \citenamefont {Kain}, \citenamefont {Ding},
  \citenamefont {Shearer}, \citenamefont {Hamers}, \citenamefont {Jin},\ and\
  \citenamefont {Wright}}]{Czech_Wright_2015}%
  \BibitemOpen
  \bibfield  {author} {\bibinfo {author} {\bibfnamefont {K.~J.}\ \bibnamefont
  {Czech}}, \bibinfo {author} {\bibfnamefont {B.~J.}\ \bibnamefont {Thompson}},
  \bibinfo {author} {\bibfnamefont {S.}~\bibnamefont {Kain}}, \bibinfo {author}
  {\bibfnamefont {Q.}~\bibnamefont {Ding}}, \bibinfo {author} {\bibfnamefont
  {M.~J.}\ \bibnamefont {Shearer}}, \bibinfo {author} {\bibfnamefont {R.~J.}\
  \bibnamefont {Hamers}}, \bibinfo {author} {\bibfnamefont {S.}~\bibnamefont
  {Jin}}, \ and\ \bibinfo {author} {\bibfnamefont {J.~C.}\ \bibnamefont
  {Wright}},\ }\href {\doibase 10.1021/acsnano.5b05198} {\bibfield  {journal}
  {\bibinfo  {journal} {{ACS} Nano}\ }\textbf {\bibinfo {volume} {9}},\
  \bibinfo {pages} {12146} (\bibinfo {year} {2015})}\BibitemShut {NoStop}%
\bibitem [{\citenamefont {Morrow}, \citenamefont {Kohler},\ and\ \citenamefont
  {Wright}(2017)}]{Morrow_Wright_2017}%
  \BibitemOpen
  \bibfield  {author} {\bibinfo {author} {\bibfnamefont {D.~J.}\ \bibnamefont
  {Morrow}}, \bibinfo {author} {\bibfnamefont {D.~D.}\ \bibnamefont {Kohler}},
  \ and\ \bibinfo {author} {\bibfnamefont {J.~C.}\ \bibnamefont {Wright}},\
  }\href {\doibase 10.1103/physreva.96.063835} {\bibfield  {journal} {\bibinfo
  {journal} {Physical Review A}\ }\textbf {\bibinfo {volume} {96}} (\bibinfo
  {year} {2017}),\ 10.1103/physreva.96.063835}\BibitemShut {NoStop}%
\bibitem [{\citenamefont {Handali}\ \emph {et~al.}(2018)\citenamefont
  {Handali}, \citenamefont {Sunden}, \citenamefont {Kaufman},\ and\
  \citenamefont {Wright}}]{Handali_Wright_2018}%
  \BibitemOpen
  \bibfield  {author} {\bibinfo {author} {\bibfnamefont {J.~D.}\ \bibnamefont
  {Handali}}, \bibinfo {author} {\bibfnamefont {K.~F.}\ \bibnamefont {Sunden}},
  \bibinfo {author} {\bibfnamefont {E.~M.}\ \bibnamefont {Kaufman}}, \ and\
  \bibinfo {author} {\bibfnamefont {J.~C.}\ \bibnamefont {Wright}},\ }\href
  {\doibase 10.1016/j.chemphys.2018.05.023} {\bibfield  {journal} {\bibinfo
  {journal} {Chemical Physics}\ } (\bibinfo {year} {2018}),\
  10.1016/j.chemphys.2018.05.023}\BibitemShut {NoStop}%
\bibitem [{\citenamefont {Honold}\ \emph {et~al.}(1988)\citenamefont {Honold},
  \citenamefont {Schultheis}, \citenamefont {Kuhl},\ and\ \citenamefont
  {Tu}}]{Honold_Tu_1988}%
  \BibitemOpen
  \bibfield  {author} {\bibinfo {author} {\bibfnamefont {A.}~\bibnamefont
  {Honold}}, \bibinfo {author} {\bibfnamefont {L.}~\bibnamefont {Schultheis}},
  \bibinfo {author} {\bibfnamefont {J.}~\bibnamefont {Kuhl}}, \ and\ \bibinfo
  {author} {\bibfnamefont {C.~W.}\ \bibnamefont {Tu}},\ }\href {\doibase
  10.1063/1.99549} {\bibfield  {journal} {\bibinfo  {journal} {Applied Physics
  Letters}\ }\textbf {\bibinfo {volume} {52}},\ \bibinfo {pages} {2105}
  (\bibinfo {year} {1988})}\BibitemShut {NoStop}%
\bibitem [{\citenamefont {Laskar}\ \emph {et~al.}(2013)\citenamefont {Laskar},
  \citenamefont {Ma}, \citenamefont {Kannappan}, \citenamefont {Park},
  \citenamefont {Krishnamoorthy}, \citenamefont {Nath}, \citenamefont {Lu},
  \citenamefont {Wu},\ and\ \citenamefont {Rajan}}]{Laskar_Rajan_2013}%
  \BibitemOpen
  \bibfield  {author} {\bibinfo {author} {\bibfnamefont {M.~R.}\ \bibnamefont
  {Laskar}}, \bibinfo {author} {\bibfnamefont {L.}~\bibnamefont {Ma}}, \bibinfo
  {author} {\bibfnamefont {S.}~\bibnamefont {Kannappan}}, \bibinfo {author}
  {\bibfnamefont {P.~S.}\ \bibnamefont {Park}}, \bibinfo {author}
  {\bibfnamefont {S.}~\bibnamefont {Krishnamoorthy}}, \bibinfo {author}
  {\bibfnamefont {D.~N.}\ \bibnamefont {Nath}}, \bibinfo {author}
  {\bibfnamefont {W.}~\bibnamefont {Lu}}, \bibinfo {author} {\bibfnamefont
  {Y.}~\bibnamefont {Wu}}, \ and\ \bibinfo {author} {\bibfnamefont
  {S.}~\bibnamefont {Rajan}},\ }\href {\doibase 10.1063/1.4811410} {\bibfield
  {journal} {\bibinfo  {journal} {Applied Physics Letters}\ }\textbf {\bibinfo
  {volume} {102}},\ \bibinfo {pages} {252108} (\bibinfo {year}
  {2013})}\BibitemShut {NoStop}%
\bibitem [{\citenamefont {Morrow}\ \emph {et~al.}(2018)\citenamefont {Morrow},
  \citenamefont {Kohler}, \citenamefont {Czech},\ and\ \citenamefont
  {Wright}}]{OSF}%
  \BibitemOpen
  \bibfield  {author} {\bibinfo {author} {\bibfnamefont {D.}~\bibnamefont
  {Morrow}}, \bibinfo {author} {\bibfnamefont {D.}~\bibnamefont {Kohler}},
  \bibinfo {author} {\bibfnamefont {K.}~\bibnamefont {Czech}}, \ and\ \bibinfo
  {author} {\bibfnamefont {J.}~\bibnamefont {Wright}},\ }\href {\doibase
  10.17605/osf.io/2wf6g} {\  (\bibinfo {year} {2018}),\
  10.17605/osf.io/2wf6g}\BibitemShut {NoStop}%
\bibitem [{\citenamefont {Thompson}\ \emph
  {et~al.}(2018{\natexlab{a}})\citenamefont {Thompson}, \citenamefont {Sunden},
  \citenamefont {Morrow},\ and\ \citenamefont {Neff-Mallon}}]{PyCMDS}%
  \BibitemOpen
  \bibfield  {author} {\bibinfo {author} {\bibfnamefont {B.~J.}\ \bibnamefont
  {Thompson}}, \bibinfo {author} {\bibfnamefont {K.~F.}\ \bibnamefont
  {Sunden}}, \bibinfo {author} {\bibfnamefont {D.~J.}\ \bibnamefont {Morrow}},
  \ and\ \bibinfo {author} {\bibfnamefont {N.~A.}\ \bibnamefont
  {Neff-Mallon}},\ }\href {\doibase 10.5281/zenodo.1198911} {\enquote {\bibinfo
  {title} {{PyCMDS}},}\ } (\bibinfo {year} {2018}{\natexlab{a}})\BibitemShut
  {NoStop}%
\bibitem [{\citenamefont {Thompson}\ \emph
  {et~al.}(2018{\natexlab{b}})\citenamefont {Thompson}, \citenamefont {Sunden},
  \citenamefont {Morrow}, \citenamefont {Neff-Mallon}, \citenamefont {Czech},
  \citenamefont {Kohler},\ and\ \citenamefont {Swedin}}]{WrightTools}%
  \BibitemOpen
  \bibfield  {author} {\bibinfo {author} {\bibfnamefont {B.~J.}\ \bibnamefont
  {Thompson}}, \bibinfo {author} {\bibfnamefont {K.~F.}\ \bibnamefont
  {Sunden}}, \bibinfo {author} {\bibfnamefont {D.~J.}\ \bibnamefont {Morrow}},
  \bibinfo {author} {\bibfnamefont {N.~A.}\ \bibnamefont {Neff-Mallon}},
  \bibinfo {author} {\bibfnamefont {K.~J.}\ \bibnamefont {Czech}}, \bibinfo
  {author} {\bibfnamefont {D.~D.}\ \bibnamefont {Kohler}}, \ and\ \bibinfo
  {author} {\bibfnamefont {R.}~\bibnamefont {Swedin}},\ }\href {\doibase
  10.5281/zenodo.1198905} {\enquote {\bibinfo {title} {{WrightTools}},}\ }
  (\bibinfo {year} {2018}{\natexlab{b}})\BibitemShut {NoStop}%
\bibitem [{\citenamefont {Jones}, \citenamefont {Oliphant},\ and\ \citenamefont
  {Peterson}(2001)}]{Jones_2001}%
  \BibitemOpen
  \bibfield  {author} {\bibinfo {author} {\bibfnamefont {E.}~\bibnamefont
  {Jones}}, \bibinfo {author} {\bibfnamefont {T.}~\bibnamefont {Oliphant}}, \
  and\ \bibinfo {author} {\bibfnamefont {P.}~\bibnamefont {Peterson}},\ }\href
  {http://www.scipy.org/} {\enquote {\bibinfo {title} {{SciPy}: Open source
  scientific tools for {Python}},}\ } (\bibinfo {year} {2001}),\ \bibinfo
  {note} {[Online; accessed 2017-09-28]}\BibitemShut {NoStop}%
\bibitem [{\citenamefont {van~der Walt}, \citenamefont {Colbert},\ and\
  \citenamefont {Varoquaux}(2011)}]{vanderWalt_Varoquaux_2011}%
  \BibitemOpen
  \bibfield  {author} {\bibinfo {author} {\bibfnamefont {S.}~\bibnamefont
  {van~der Walt}}, \bibinfo {author} {\bibfnamefont {S.~C.}\ \bibnamefont
  {Colbert}}, \ and\ \bibinfo {author} {\bibfnamefont {G.}~\bibnamefont
  {Varoquaux}},\ }\href {\doibase 10.1109/mcse.2011.37} {\bibfield  {journal}
  {\bibinfo  {journal} {Computing in Science {\&} Engineering}\ }\textbf
  {\bibinfo {volume} {13}},\ \bibinfo {pages} {22} (\bibinfo {year}
  {2011})}\BibitemShut {NoStop}%
\bibitem [{\citenamefont {Hunter}(2007)}]{Hunter_2007}%
  \BibitemOpen
  \bibfield  {author} {\bibinfo {author} {\bibfnamefont {J.~D.}\ \bibnamefont
  {Hunter}},\ }\href {\doibase 10.1109/mcse.2007.55} {\bibfield  {journal}
  {\bibinfo  {journal} {Computing in Science {\&} Engineering}\ }\textbf
  {\bibinfo {volume} {9}},\ \bibinfo {pages} {90} (\bibinfo {year}
  {2007})}\BibitemShut {NoStop}%
\bibitem [{\citenamefont {Zhang}\ \emph {et~al.}(2015)\citenamefont {Zhang},
  \citenamefont {Dong}, \citenamefont {McEvoy}, \citenamefont {O'Brien},
  \citenamefont {Winters}, \citenamefont {Berner}, \citenamefont {Yim},
  \citenamefont {Li}, \citenamefont {Zhang}, \citenamefont {Chen},
  \citenamefont {Zhang}, \citenamefont {Duesberg},\ and\ \citenamefont
  {Wang}}]{Zhang_Wang_2015}%
  \BibitemOpen
  \bibfield  {author} {\bibinfo {author} {\bibfnamefont {S.}~\bibnamefont
  {Zhang}}, \bibinfo {author} {\bibfnamefont {N.}~\bibnamefont {Dong}},
  \bibinfo {author} {\bibfnamefont {N.}~\bibnamefont {McEvoy}}, \bibinfo
  {author} {\bibfnamefont {M.}~\bibnamefont {O'Brien}}, \bibinfo {author}
  {\bibfnamefont {S.}~\bibnamefont {Winters}}, \bibinfo {author} {\bibfnamefont
  {N.~C.}\ \bibnamefont {Berner}}, \bibinfo {author} {\bibfnamefont
  {C.}~\bibnamefont {Yim}}, \bibinfo {author} {\bibfnamefont {Y.}~\bibnamefont
  {Li}}, \bibinfo {author} {\bibfnamefont {X.}~\bibnamefont {Zhang}}, \bibinfo
  {author} {\bibfnamefont {Z.}~\bibnamefont {Chen}}, \bibinfo {author}
  {\bibfnamefont {L.}~\bibnamefont {Zhang}}, \bibinfo {author} {\bibfnamefont
  {G.~S.}\ \bibnamefont {Duesberg}}, \ and\ \bibinfo {author} {\bibfnamefont
  {J.}~\bibnamefont {Wang}},\ }\href {\doibase 10.1021/acsnano.5b03480}
  {\bibfield  {journal} {\bibinfo  {journal} {{ACS} Nano}\ }\textbf {\bibinfo
  {volume} {9}},\ \bibinfo {pages} {7142} (\bibinfo {year} {2015})}\BibitemShut
  {NoStop}%
\bibitem [{\citenamefont {McIntyre}\ and\ \citenamefont
  {Aspnes}(1971)}]{McIntyre_Aspens_1971}%
  \BibitemOpen
  \bibfield  {author} {\bibinfo {author} {\bibfnamefont {J.}~\bibnamefont
  {McIntyre}}\ and\ \bibinfo {author} {\bibfnamefont {D.}~\bibnamefont
  {Aspnes}},\ }\href {\doibase 10.1016/0039-6028(71)90272-x} {\bibfield
  {journal} {\bibinfo  {journal} {Surface Science}\ }\textbf {\bibinfo {volume}
  {24}},\ \bibinfo {pages} {417} (\bibinfo {year} {1971})}\BibitemShut
  {NoStop}%
\bibitem [{\citenamefont {Mak}\ \emph {et~al.}(2008)\citenamefont {Mak},
  \citenamefont {Sfeir}, \citenamefont {Wu}, \citenamefont {Lui}, \citenamefont
  {Misewich},\ and\ \citenamefont {Heinz}}]{Mak_Heinz_2008}%
  \BibitemOpen
  \bibfield  {author} {\bibinfo {author} {\bibfnamefont {K.~F.}\ \bibnamefont
  {Mak}}, \bibinfo {author} {\bibfnamefont {M.~Y.}\ \bibnamefont {Sfeir}},
  \bibinfo {author} {\bibfnamefont {Y.}~\bibnamefont {Wu}}, \bibinfo {author}
  {\bibfnamefont {C.~H.}\ \bibnamefont {Lui}}, \bibinfo {author} {\bibfnamefont
  {J.~A.}\ \bibnamefont {Misewich}}, \ and\ \bibinfo {author} {\bibfnamefont
  {T.~F.}\ \bibnamefont {Heinz}},\ }\href {\doibase
  10.1103/physrevlett.101.196405} {\bibfield  {journal} {\bibinfo  {journal}
  {Physical Review Letters}\ }\textbf {\bibinfo {volume} {101}} (\bibinfo
  {year} {2008}),\ 10.1103/physrevlett.101.196405}\BibitemShut {NoStop}%
\bibitem [{\citenamefont {Sipe}\ and\ \citenamefont
  {Shkrebtii}(2000)}]{Sipe_Shkrebtii_2000}%
  \BibitemOpen
  \bibfield  {author} {\bibinfo {author} {\bibfnamefont {J.~E.}\ \bibnamefont
  {Sipe}}\ and\ \bibinfo {author} {\bibfnamefont {A.~I.}\ \bibnamefont
  {Shkrebtii}},\ }\href {\doibase 10.1103/physrevb.61.5337} {\bibfield
  {journal} {\bibinfo  {journal} {Physical Review B}\ }\textbf {\bibinfo
  {volume} {61}},\ \bibinfo {pages} {5337} (\bibinfo {year}
  {2000})}\BibitemShut {NoStop}%
\bibitem [{\citenamefont {Axt}\ and\ \citenamefont
  {Mukamel}(1998)}]{Axt_Mukamel_1998}%
  \BibitemOpen
  \bibfield  {author} {\bibinfo {author} {\bibfnamefont {V.~M.}\ \bibnamefont
  {Axt}}\ and\ \bibinfo {author} {\bibfnamefont {S.}~\bibnamefont {Mukamel}},\
  }\href {\doibase 10.1103/revmodphys.70.145} {\bibfield  {journal} {\bibinfo
  {journal} {Reviews of Modern Physics}\ }\textbf {\bibinfo {volume} {70}},\
  \bibinfo {pages} {145} (\bibinfo {year} {1998})}\BibitemShut {NoStop}%
\bibitem [{\citenamefont {Peyghambarian}, \citenamefont {Koch},\ and\
  \citenamefont {Mysyrowicz}(1993)}]{Peychambarian_Mysyrowicz_1993}%
  \BibitemOpen
  \bibfield  {author} {\bibinfo {author} {\bibfnamefont {N.}~\bibnamefont
  {Peyghambarian}}, \bibinfo {author} {\bibfnamefont {S.~W.}\ \bibnamefont
  {Koch}}, \ and\ \bibinfo {author} {\bibfnamefont {A.}~\bibnamefont
  {Mysyrowicz}},\ }\href@noop {} {\emph {\bibinfo {title} {Introduction to
  Semiconductor Optics}}}\ (\bibinfo  {publisher} {Prentice Hall},\ \bibinfo
  {year} {1993})\BibitemShut {NoStop}%
\bibitem [{\citenamefont {Dresselhaus}\ \emph {et~al.}(2018)\citenamefont
  {Dresselhaus}, \citenamefont {Dresselhaus}, \citenamefont {Cronin},\ and\
  \citenamefont {Filho}}]{Dresselhaus_2018}%
  \BibitemOpen
  \bibfield  {author} {\bibinfo {author} {\bibfnamefont {M.}~\bibnamefont
  {Dresselhaus}}, \bibinfo {author} {\bibfnamefont {G.}~\bibnamefont
  {Dresselhaus}}, \bibinfo {author} {\bibfnamefont {S.}~\bibnamefont {Cronin}},
  \ and\ \bibinfo {author} {\bibfnamefont {A.~G.~S.}\ \bibnamefont {Filho}},\
  }\href {\doibase 10.1007/978-3-662-55922-2} {\emph {\bibinfo {title} {Solid
  State Properties}}}\ (\bibinfo  {publisher} {Springer Berlin Heidelberg},\
  \bibinfo {year} {2018})\BibitemShut {NoStop}%
\bibitem [{\citenamefont {Padilha}\ \emph {et~al.}(2014)\citenamefont
  {Padilha}, \citenamefont {Peelaers}, \citenamefont {Janotti},\ and\
  \citenamefont {de~Walle}}]{Padilha_VandeWalle_2014}%
  \BibitemOpen
  \bibfield  {author} {\bibinfo {author} {\bibfnamefont {J.~E.}\ \bibnamefont
  {Padilha}}, \bibinfo {author} {\bibfnamefont {H.}~\bibnamefont {Peelaers}},
  \bibinfo {author} {\bibfnamefont {A.}~\bibnamefont {Janotti}}, \ and\
  \bibinfo {author} {\bibfnamefont {C.~G.~V.}\ \bibnamefont {de~Walle}},\
  }\href {\doibase 10.1103/physrevb.90.205420} {\bibfield  {journal} {\bibinfo
  {journal} {Physical Review B}\ }\textbf {\bibinfo {volume} {90}} (\bibinfo
  {year} {2014}),\ 10.1103/physrevb.90.205420}\BibitemShut {NoStop}%
\bibitem [{Note1()}]{Note1}%
  \BibitemOpen
  \bibinfo {note} {\protect \autoref {E:chi1}, and all further theory
  developed, neglect indirect transitions. We find this a reasonable assumption
  since our multidimensional spectrum exhibited no cross-peaks between the
  K-point features (A and B) and the C band.}\BibitemShut {Stop}%
\bibitem [{\citenamefont {Kuzmenko}(2005)}]{Kuzmenko_2005}%
  \BibitemOpen
  \bibfield  {author} {\bibinfo {author} {\bibfnamefont {A.~B.}\ \bibnamefont
  {Kuzmenko}},\ }\href {\doibase 10.1063/1.1979470} {\bibfield  {journal}
  {\bibinfo  {journal} {Review of Scientific Instruments}\ }\textbf {\bibinfo
  {volume} {76}},\ \bibinfo {pages} {083108} (\bibinfo {year}
  {2005})}\BibitemShut {NoStop}%
\bibitem [{\citenamefont {Sie}\ \emph {et~al.}(2015)\citenamefont {Sie},
  \citenamefont {Frenzel}, \citenamefont {Lee}, \citenamefont {Kong},\ and\
  \citenamefont {Gedik}}]{Sie_Gedik_2015}%
  \BibitemOpen
  \bibfield  {author} {\bibinfo {author} {\bibfnamefont {E.~J.}\ \bibnamefont
  {Sie}}, \bibinfo {author} {\bibfnamefont {A.~J.}\ \bibnamefont {Frenzel}},
  \bibinfo {author} {\bibfnamefont {Y.-H.}\ \bibnamefont {Lee}}, \bibinfo
  {author} {\bibfnamefont {J.}~\bibnamefont {Kong}}, \ and\ \bibinfo {author}
  {\bibfnamefont {N.}~\bibnamefont {Gedik}},\ }\href {\doibase
  10.1103/physrevb.92.125417} {\bibfield  {journal} {\bibinfo  {journal}
  {Physical Review B}\ }\textbf {\bibinfo {volume} {92}} (\bibinfo {year}
  {2015}),\ 10.1103/physrevb.92.125417}\BibitemShut {NoStop}%
\bibitem [{\citenamefont {Roessler}(1965)}]{Roessler_1965}%
  \BibitemOpen
  \bibfield  {author} {\bibinfo {author} {\bibfnamefont {D.~M.}\ \bibnamefont
  {Roessler}},\ }\href {\doibase 10.1088/0508-3443/16/8/310} {\bibfield
  {journal} {\bibinfo  {journal} {British Journal of Applied Physics}\ }\textbf
  {\bibinfo {volume} {16}},\ \bibinfo {pages} {1119} (\bibinfo {year}
  {1965})}\BibitemShut {NoStop}%
\end{thebibliography}%

\end{document}